\documentstyle{amsart}
\newcommand{\R}{{\Bbb R}}
\newcommand{\Z}{{\Bbb Z}}

\newcommand{\Hb}{{\Bbb H}}
\newcommand{\C}{{\Bbb C}}

\newcommand{\T}{{\cal T}}
\newcommand{\Hh}{{\cal H}}

\newcommand{\N}{{\cal N}}

\newcommand{\E}{{\cal E}}

\newcommand{\hol}{\operatorname{hol}}
\newcommand{\ind}{\operatorname{ind}}

\newcommand{\itt}{\operatorname{it}}

\newcommand{\red}{\operatorname{red}}

\newcommand{\Spin}{\operatorname{Spin}}

\newcommand{\Tr}{\operatorname{Tr}}

\newcommand{\spann}{\operatorname{span}}

\newcommand{\half}{\frac{1}{2}}

\newcommand{\sa}{\sigma_A}

\newcommand{\pis}{\Pi_\Sigma}

\newsymbol\upharpoonright 1318
\newsymbol\dhr 1317

\theoremstyle{plain}

\theoremstyle{definition}

\theoremstyle{remark}

\title{Index and Dynamics of quantized contact transformations }
\author{Steven Zelditch*}
\date{September 1995}
\thanks{ *Partially supported by NSF grant \#DMS-9404637.}
\address{Johns Hopkins University, Baltimore, Maryland  21218}

\begin{document}
\maketitle
\addtolength{\baselineskip}{1pt}
\begin{abstract}
Quantized contact transformations are Toeplitz operators over a contact manifold $(X,\alpha)$
of the form $U_{\chi} = \Pi A \chi \Pi$, where $\Pi : H^2(X) \rightarrow L^2(X)$ is a
Szego projector, where $\chi$ is a contact transformation and where $A$ is a 
pseudodifferential operator over $X$.  They provide a flexible alternative to the Kahler
quantization of symplectic maps, and encompass many of the examples in the
physics literature, e.g. quantized cat maps and kicked rotors.  The index problem is to
determine $ind(U_{\chi})$ when the principal symbol is unitary, or equivalently to
determine whether $A$ can be chosen so that $U_{\chi}$ is unitary.  We show that the
answer is yes in the case of quantized symplectic torus automorphisms $g$---by
showing that $U_g$ duplicates the classical transformation laws on theta functions.
Using the Cauchy-Szego kernel on the Heisenberg group, we calculate the traces on 
theta functions of each degree N.  We also study the quantum dynamics generated by
a general q.c.t. $U_{\chi}$, i.e. the quasi-classical asymptotics of the eigenvalues and
eigenfunctions under various ergodicity and mixing hypotheses on $\chi.$  Our principal
results are proofs of equidistribution of eigenfunctions $\phi_{Nj}$ and weak mixing
properties of matrix elements $(B\phi_{Ni}, \phi_{Nj})$ for quantizations of mixing
symplectic maps.

\end{abstract} 

\setcounter{section}{-1}
\section{Introduction}

The problem of quantizing symplectic maps and of analyzing the dynamics
of the quantum system is a very basic one in mathematical physics, and has
been studied extensively, by both mathematicians and physicists, from many
 different points of view.   The present article is concerned with one such
quantization method, that of {\it Toeplitz quantization}, and with one particular
viewpoint towards the ergodicity and mixing properties of the quantized maps,
that of {\it quantized Gelfand-Naimark-Segal systems}  in the sense of [Z.1].   We will describe a method
of quantizing contact transformations of a contact manifold $(X,\alpha)$ with
periodic contact flow as unitary operators on an associated Hardy space $H^2(X)$ ,
and prove a number of results on the index and dynamics of the quantized contact
transformations.  The method, essentially a unitarized version of Boutet de
Monvel's Toeplitz quantization [B][B.G], is closely related to the geometric quantization
of symplectic maps on Kahler manifolds and produces  new examples of quantized
GNS systems.  The  quantum ergodicity theorems follow in part from the general
results of [Z.1], but also include some sharper ergodicity and mixing theorems
analogous to those of [Z.2,3] in the case of wave groups.

To illustrate the method and ergodicity results 
we will also study in detail the Toeplitz  quantization of  symplectic
torus automorphisms ( `cat maps') (\S 5),  undoubtedly the most popular of maps to undergo quantization--
see [A.d'P.W][B.N.S][H.B][dB.B][d'E.G.I][K.P][K][Ke][We] for just a few among the many treatments.
As the reader is surely aware, {\it quantization} is not a uniquely defined process
and it is not apriori clear how the plethora of  quantizations defined  in these
articles are  related to each other or to the quantization presented here.  In fact, although
it is not quite  obvious, all but that of [B.N.S] are  equivalent to the Toeplitz quantization
presented here.    We will describe the relations between them more precisely
 at the end of the introduction.

  What is more, it  will be proved in \S 5 that the Toeplitz quantization of $SL(2,\Z)$
 reproduces what must be the quantization of most ancient vintage----namely, the
Hermite-Jacobi action of $SL(2,\Z)$ (or more precisely its theta-subgroup
 $SL_{\theta}(2, \Z)$) on 
spaces $\Theta_N$ of theta functions of any degree $N$ (cf.  [Herm], [Kloo] or, for a modern treatment, [K][K.P]).
We construct this action by lifting $g \in SL_{\theta}(2, \Z)$ to a contact transformation
$\chi_g$ of $N_{\R} / N_{\Z}$, the quotient of the Heisenberg group by its integral subgroup,
and compressing the latter to the spaces $\Theta_N$.
This connection develops the long chain of links between theta functions and
 harmonic analysis on the Heisenberg group  (e.g. [A][M]), and is perhaps
of independent interest.   For one thing, it gives a framework for analysing asymptotic
properties of theta functions in the semi-classical (large $N$) limit.  It may also be
 used, together with the explicit formula
of the Cauchy-Szego kernel on the Heisenberg group,  to give a Selberg-type 
 trace formula for the trace of an element $g \in SL_{\theta}(2, \Z)$
 acting on the space of  theta functions of degree N (\S 6).   

It should also be mentioned that the quantization of $SL_{\theta}(2, \Z)$ as unitary operators
$U_{g,N}$ on $\Theta_N$ is just a concrete realization of the metaplectic representation of the
finite metaplectic group
$Mp(2, \Z/N)$.  That is, the Toeplitz-quantization of an element $g \in SL_{\theta}(2, \Z)$
is equivalent to reducing it mod $N$,
and then applying the  metaplectic representation $\mu_N$ of $SL(2, \Z/N).$  Hence the
trace formula alluded to above is giving the characters  of the finite metaplectic representations.
We further mention that when $N = p^k$ is a power of a prime,  $U_{g,N}$ may be described in terms
of the metaplectic representation over the field of p-adic numbers, indeed 
as the quantization of $g$ viewed as a symplectic map on the p-adic torus.  We will not
develop this point of view here,  but we hope it will eventually help clarify the number-theoretic
 aspects of the spectral theory of $U_{g,N}$ (cf. [Ke][d'E.G.I] ).

Although our main aim in this article is to discuss the quantum dynamics of Toeplitz-
quantized maps, we would also like to recall that there is an interesting index problem
associated to them.   Namely, the Toeplitz-quantization of a symplectic or contact
 map $\chi$ will be an operator
$U_{\chi}$ which is unitary modulo finite rank operators.  It therefore has an index, which
depends only on $\chi$ and on the principal symbol of $U_{\chi}$.   The problem of calculating
this index $ind(\chi)$ was raised in [Wei] in the closely related 
 context of Fourier Integral operators but it does
not seem to have been calculated before in any example.   Hence it may be of interest to observe
 that the index $ind(\chi_g)$ of $g \in SL(2,\Z)$  is always zero, as follows 
 the unitarity of the Hermite-Jacobi `transformation laws'.   This vanishing of 
the index has a very simple alternative explanation, so it is not clear how
generally to expect the index to vanish
 (see the Remarks at the end of \S 5).

This article will presume a degree of familiarity with the machinery of Toeplitz operators
as presented in the book of Boutet-de-Monvel and Guillemin [B.G].  This machinery 
involves 
some language and ideas from  symplectic geometry, microlocal analysis, 
 several complex variables, CR functions
and from the representation theory of the Heisenberg and metaplectic groups.   We hope
that the explicit calculations of symbols, quantizations, traces and so on in the case of
the symplectic torus automorphisms
 will provide  elementary examples of how this machinery works, in a form accessible to
those studying quantum maps from other points of view.   In an obvious sense,
which should be clearer by the end of \S 5-6, the cat maps
are among the  basic linear models for the general theory.   

We will also assume some familiarity with quantum dynamics, especially from the
semi-classical viewpoint.   This is actually a rather broad assumption, since there are
many different approaches to quantum dynamics.  With the aim of clarifying the relation
between our set-up, methods and results with those of other articles  
on the dynamics of quantum maps, we
 end this introduction with a rapid comparison  to the works cited above.
\bigskip

\noindent{\bf 0.1 Comparison to prior articles}
\bigskip

First, let us compare quantization methods.   Besides the Toeplitz method of
quantizing a symplectic map on a compact symplectic manifold,  which requires the map
to lift as a contact map of the `prequantum circle bundle,' 
the only general method is that of geometric quantization.   Traditionally, this is a 
method only of quantizing symplectic manifolds and observables; but in the last few
years it has been extended to include a variety of symplectic maps.  In particular,
motivated by the needs of  topological field theory,  there are many articles using
the method of Kahler quantization to quantize elements of  $SL(2,\Z)$.  By Kahler
quantization we mean geometric quantization on a Kahler manifold in the presence
of a complex polarization.  This is the method used in [A.dP.W][We], among many other
places,  and discussed in the book of Atiyah [At].  As in the Toeplitz construction,
the symplectic torus automorphisms are quantized as translation operators on
theta functions.  However,  such translations change the complex structure and so
do not preserve a fixed space of holomorphic theta functions.  In the language of
geometric quantization, one has to define a BKS (Blattner-Kostant-Sternberg)
pairing between the different complex polarizations to return to the original space.
It is at this point that the Toeplitz and Kahler methods differ:
 In the Toeplitz method, one
uses  orthogonal projection back to the original space (times a normalizing factor)
while in the Kahler method, one uses a parallel  translation along the moduli space of
complex structures on the torus.  In the case of torus automorphisms, both  methods
produce  the classical transformation laws for theta functions (in the Kahler case,
this was  pointed out by  Weitsman (loc.cit.).  Hence the Toeplitz
and Kahler methods are
equivalent in this case. They are surely equivalent in much greater generality, but to
the author's knowledge this has never been studied systematically.

 The other quantizations of the cat maps [H.B][Kea][dE.G.I][dB.B][B.N.S] are based 
 (implicitly or explicitly) on the  
special representation
theory of the Heisenberg and metaplectic groups.  This is also true in the many physics
articles on other quantum maps such as  kicked rotors and tops.  It is the author's
impression that the methods of geometric and Kahler quantization are seldom used
in the semi-classical physics literature, wherein quantization seems to be equated
with canonical quantization (i.e. with representation theory of the Heisenberg
group).  It may therefore be useful to point out that the
Toeplitz method gives equivalent quantizations to `Weyl'
or 'canonical quantization',
not only for the cat maps but also for all other symplectic maps mentioned above.

Now let us turn to the  comparison of dynamical notions such as ergodicity, mixing, K
and so on. 

These notions are often left undefined in the semi-classical literature, since the
main problem there is to determine the impact of dynamical properties of the classical
limit on the spectral data of the quantum system.   However,  one can also introduce
intrinsic notions of quantum ergodicity, mixing, complete integrability (and so on)
which capture the behaviour of quantizations of classically ergodic (etc.) systems. 
The definitions used in this paper are of this kind;  they are  based on [Z.1,2] (see
also [Su]) and will be reviewed in \S 2.

In contrast,  there are the definitions of ergodicity, mixing (etc.) in the theory
of   $C^*$- or $W^*$-dynamical systems.  These are more analogous to the classical
notions and are applied to open or infinite quantum systems.   In this framework,
 a quantum dynamical system is defined by a C* or W* algebra ${\cal A}$ of observables,
together with an action $\alpha: G \rightarrow Aut({\cal A})$ of a group $G$ by
automorphisms of ${\cal A}$.  The system
$({\cal A}, G, \alpha)$
is generally covariantly represented on a Hilbert space ${\cal H}$,
so that $\alpha_g(A) = U_g^*AU_g$ of ${\cal A}$, 
with $U_g$ a unitary representation of $G$ on ${\cal H}$.  Dynamical notions are  
non-commutative analogues, often at the von-Neumann algebra ($W^*$-) level,
 of the usual notions for abelian systems.   In particular, the spectra of mixing systems
must be  continuous.  For some recent references in the mathematical
physics literature, see [B][J.P][Th].

As mentioned above,
our interest   is in the semi-classical aspects of quantum dynamics:
The quantum systems studied in this paper
 will have discrete spectra and the ergodicity and mixing properties
will be reflected (by definition) in the asymptotics of the eigenfunctions and the eigenvalues.
To clarify the relation between this point of view and that of the  C*-dynamical point of view,
we will also state definitions in terms of the relevant C* algebras and their automorphism
groups.  It is hoped that this approach will also clarify the nature of the dynamical properties
at issue in the semi-classical literature.

Let us contrast the two kinds of dynamical notions in the example of the cat maps, using
 the articles   [B] [B.N.S] [N.T1,2][Th] to represent the C* and W* approach.
 In these artricles, the cat maps
 are quantized as automorphisms of the
rotation algebras ${\cal M}_{\theta}$ (the non-commutative torus), and have precisely
one invariant state.  The GNS representation with respect to this state determines a
covariant representation of this system by translations by the classical cat map on
functions on the torus.  In  their
words, this gives a ``radically different" quantization from the semi-classical one, in that the   
 quantized cat maps have the same multiple Lesbesque spectrum, hence the same
mixing properties, as the classical maps. 

 The relation of this to the Toeplitz (or other semi-classical) quantizations is as follows: first, 
in the semi-classical
quantizations, the Planck constant $\theta$ varies only over the rational values $
\frac{1}{N}$, corresponding to the space $\Theta_N$ of theta functions of degree N. 
The finite Heisenberg group $Heis(\Z /N)$ acts irreducibly on this space and its
group algebra
$\C[Heis(\Z /N)]$ defines the relevant C* algebra.  This algebra is not the rotation algebra
${\cal M}_{\frac{1}{N}}$ but is rather the quotient ${\cal M}_{\frac{1}{N}}/{\cal Z}_N$
by its center ${\cal Z}_N$.  The elements of $SL(2,Z)$ define automorphisms of
${\cal M}_{\frac{1}{N}}$ which (under a parity assumption) preserve the center.  Hence
they also define automorphisms of the quotient algebra.  The quotient automorphisms
are the ones studied in the semi-classical literature.  Unlike the automorphisms of the
full rotation algebras, the quotient ones have discrete spectra and
  many invariant states, and hence are not ergodic in the C* sense.  However, 
 they are quantum ergodic in the semi-classical sense whenever the classical cat map
is ergodic.   Finally, we note that
the quantized cap map systems in the sense of [B.N.S]  are also
quantized GNS systems in the sense of [Z.1], and are trivially quantum ergodic because the
only invariant state is the unique tracial state.  Hence
 they do not have distinct classical limits in the sense of this paper.
From our point of view, therefore,
the quantizations in [B][B.N.S][NT1,2]  appear essentially as classical
dynamical systems, albeit involving non-commutative algebras. 
  See \S 5 for a more complete discussion.

The organization of this article is as follows:
\bigskip

\begin{tabular}{ll} \S 0 & Introduction \\
\S 1 & Statement of results \\
\S 2 & Background \\
\S 3 & Symplectic spinors and proof of the unitarization lemma \\
\S 4 & Quantum ergodicity and mixing: Proof of Theorems A,B,C \\
\S 5 & Quantized symplectic torus automorphisms: Proof of Theorem D\\
\S 6 & Trace formulae for quantized toral automorphisms. \end{tabular}
\bigskip

\subsection*{Acknowledgements} Conversations with A.Uribe on Toeplitz operators, with
A.Weinstein on the index of a contact transformation,
and with J.Weitsman on Kahler quantization and theta-functions are gratefully acknowledged. 
We have also profited from an unpublished article of V.Guillemin [G.2], which discusses
the trace formula in Theorem E for elliptic elements.  

\section{Statement of results}

In this article, the terms {\it quantum ergodicity} and {\it quantum mixing} refer
to the properties of {\it quantized abelian systems}
defined in [Z.1,2].  They will be briefly reviewed in
\S2.

 We will be concentrating on one kind of example of such quantized abelian systems.  The
setting will
consist of a   compact contact manifold $(X, \alpha)$ 
with an  periodic contact flow $\phi^t$, together with a contact
transformation 
$$\chi: X \rightarrow X\;\;\;\;\;\chi^*(\alpha) = \alpha\;\;\;\;\chi \cdot \phi^t
=\phi^t \cdot \chi$$
 commuting with $\phi^t.$  
 The $S^1$ action defined by $\phi^t$ will be assumed elliptic, so that its
isotypic spaces are finite dimensional. The map $\chi$
will be quantized as  a Toeplitz- Fourier Integral operator 
$$U_{\chi}:  H^2_{\Sigma}(X)\rightarrow   H^2_{\Sigma}(X)   $$  
acting on a  Hilbert space $H^2_{\Sigma}(X)$  of generalized CR functions
on $X$ called the Hardy space.  
The motivating example is where the symplectic quotient $({\cal O},\omega)$
is a Kahler manifold and where $(X,\alpha)$ is the principal
 $U(1)$-bundle (with connection) associated to the pre-quantum line bundle (with connection)
$(L,\nabla) \rightarrow {\cal O}$  such that $curv(\nabla)=\omega.$
Relative to the given  complex structure $J$, the quantum Hilbert spaces
are the spaces ${\cal H}_J^N$ of holomorphic sections of $L^{\otimes N}$, which are canonically
isomorphic to the spaces $H^2_{\Sigma}(N)$ of $U(1)$- equivariant CR functions on $X$, 
in the CR structure induced by $J$. For precise definitions and references, see \S2-3.

We first give some general results on the spectrum and on the 
quantum ergodicity and mixing
properties of  quantized
contact transformations.  The quantization
 $U_{\chi}$ and the orthogonal projection $\Pi_{\Sigma}$ on $ H^2_{\Sigma}$ 
will be constructed so that they   
commute with  the operator $W_t$ of translation by $\phi^t$; under this $U(1)$-action,
 $ H^2_{\Sigma}$  breaks up into finite
dimensional ``weight" spaces $ H^2_{\Sigma}(N)$ of dimensions $d_N$  
and $U_{\chi}$  breaks up into  rank $d_N$  unitary operators $U_{\chi,N}.$
Hence the quantum system decomposes into finite dimensional systems.  From
the semi-classical point of view, the focus is on the eigenvalue problems:
$$\left \{ \begin{array}{ll} U_{\chi,N} \phi_{Nj} = e^{i \theta_{Nj}}\phi_{Nj} & (\phi_{Nj}
\in {\cal H}_{\Sigma}(N)) \\ (\phi_{Ni}, \phi_{Mj}>= \delta_{MN}\delta_{ij}.
 \end{array}\right\}$$

We will prove the following statements about the eigenvalues and eigenfunctions in \S4.  The
first is a rather basic and familiar kind of eigenvalue distribution theorem, which will be stated
more precisely in \S4.
\medskip

\noindent{\bf Theorem A} \;\;{\it  With the above notation and assumptions: The
spectrum $\sigma(U_{\chi})$ is a pure point spectrum.  The following dichotomy
holds:

\noindent(i) {\it aperiodic case}\;\;If the set of periodic points of $\chi$ on the symplectic
quotient ${\cal O}$ has measure zero (w.r.t. $\mu$), then
as $N \rightarrow \infty$,
 the eigenvalues $\{e^{i\theta_{Nj}}\}$ become uniformly distributed on $S^1$;

\noindent(ii) {\it periodic case}\;\;If $\chi^p=id$ for some $p>0$ then there exists
a $\chi $-invariant Toeplitz structure $\Pi_{\Sigma}$ so that $\sigma (U_{\chi}$) 
is contained in the pth roots of unity.} 
\medskip

Next comes a series of general results on the quantum dynamics of Toeplitz systems. 
The rationale for viewing them as quantum ergodicity and mixing theorems will be
reviewed in \S2 (see also [Z.1,2] for extended discussions). 
\medskip

\noindent{\bf Theorem B}\;\;{\it With the same notation and assumptions:
Suppose that $(\phi^t, \chi)$ defines an ergodic action of
$G=S^1 \times \Z$ on $(X, \alpha \wedge (d\alpha)^{n-1})$, and let $({\cal O},\omega)$ denote
the symplectic quotient.
 Then the quantized action
$(W_t, U_{\chi,a})$  of G has the following properties: for any $\sigma \in C^{\infty}({\cal O})$

$$\lim_{N\rightarrow \infty} \frac{1}{d_N}\sum_{j=1}^{d_N}
|(\sigma\phi_{Nj},\phi_{Nj})-\bar\mu(\sigma)|^2=0.\
 \leqno({\cal EP})$$
\medskip

 $$(\forall \epsilon)(\exists \delta)
\limsup_{N \rightarrow \infty} 
 \frac{1}{d_N}\sum_{{i \not= j: }\atop
{ |e^{i\theta_{Ni}} -e^{i\theta_{Nj}}| < \delta}} |( \sigma \phi_{Ni},
\phi_{Nj} )|^2 < \epsilon \leqno({\cal EP!})$$
Here, $\mu$ is the symplectic volume measure of $({\cal O},\omega)$ and $\bar\mu(\sigma) =
\frac{1}{\mu({\cal O})}\int_{{\cal O}}a d\mu$ is the average of
 $\sigma$  on $({\cal O},d\mu)$ }

\medskip

\noindent{\bf Corollary B}
{\it  For each $N$ there is a subset $J_N \subset \{1, \dots, d_N\}$ such that:
\medskip

\noindent(a) $\lim_{N \rightarrow \infty} \frac{ \# J_N}{d_N} = 1$;\\
\noindent(b) w- $\lim_{N \rightarrow \infty, j \in J_N} |\phi_{Nj}|^2 =1$ on
the quotient ${\cal O}:= X/S^1$.  Here, w-$lim$ is the weak* limit on $C({\cal O})$. }
\medskip

\noindent{\bf Theorem C} {\it With the notations and assumptions of Theorem B:
  If the action is also weak-mixing, then in addition to ${\cal EP}, {\cal EP!}$, we have,
for any $\sigma \in  C^{\infty}({\cal O})$, 
$$(\forall \epsilon)(\exists \delta)
\limsup_{N \rightarrow \infty} 
 \frac{1}{d_N}\sum_{{i\not= j: }\atop
{ |e^{i\theta_{Ni}} - e^{i\theta_{Nj}}-e^{i\tau}| < \delta}} |( \sigma \phi_{Ni},
\phi_{Nj} )|^2 < \epsilon \leqno({\cal MP})\;\;\;\;\;\;\;
(\forall \tau  \in {\bf R})$$}
\medskip

The restriction $i\not =j$ is of course redundant unless $\tau = 0$,
in which case the statement coincides with ${\cal EP!}.$  For background on these
mixing properties see \S2 and [Z.2-3].

The third series of results concerns the special case of quantized symplectic torus
automorphisms, or quantum `cat maps' (as they are known in the physics literature).
  In this case, the phase space is the torus $\R^{2n}/\Z^{2n}$,
equipped with  the standard symplectic structure $\sum dx_i \wedge d\xi_i.$
The cat maps are 
defined by elements $g \in Sp(2n, \Z)$ (or more precisely, elements of the ``theta-group"
$Sp_{\theta}(2n, \Z)$, see \S 5). 
 As will be seen in \S5 (and as is  easy to prove) 
 these symplectic maps are "contactible": i.e.  can be lifted to
the prequantum $U(1)$- bundle $X$ as  contact transformations 
  $\chi_g$.   The resulting situation
 is very nice (and very well-studied) because
of its relation to the representation theory of the Heisenberg group: This stems from the
fact that  $X$ is the compact nil-manifold $\Hb^{\red}_n/ \bar \Gamma$ where
 $\Hb^{\red}_n\sim \R^{2n} \times S^1$is the reduced Heisenberg group and
where $\Gamma$ is  the integral lattice $\Z^{2n}\times \{1\}$.  

 The spectral
theory of the classical cat map is that of the
 the unitary translation operator $T_{\chi_g}$  by $\chi_g$  on $L^2(X)$.  Its
quantization $U_g$ will be more or less its compression to the Hardy space $H^2_{\Sigma}(X)$ 
of CR functions associated to the standard CR structure on $X$.  That is,
essentially $U_g = \Pi_{\Sigma} T_{\chi_g}\Pi_{\Sigma}$ where
  $\Pi_{\Sigma}: L^2(X) \rightarrow H^2_{\Sigma}(X)$  is the Szego projector.   (As
will be explained in \S2 and \S5, this definition has only to be adjusted by a
constant so that $U_g$ is unitary. ) The projector will often be denoted more simply
by $\Pi$ when the complex or CR structure is fixed. In a
well known way, this space of CR functions can be identified with the space of theta
functions of all degrees for the lattice $\Z^{n}$, and thus the quantized cat maps
will correspond to a sequence $U_{g,N}$ of unitary operators on the spaces of theta
functions of degree N. As mentioned above, they are of a classic vintage and appear
in  the transformation laws of theta-functions.  Equivalently, they arise in the
metaplectic representation of the finite symplectic groups $Sp(2n, \Z/N).$  
From the geometric point of view, the CR compression
plays the role of the complex polarization in Kahler quantization, with the
standard CR structure corresponding to the  choice of
complex structure $J= iI$ on $\R^{2n}/\Z^{2n}$.

 Postponing complete definitions
until \S2 and \S5, we may state our results on theta functions as follows
\medskip

\noindent{\bf Theorem D} {\it Let $g \in SL_{\theta}(2,\Z):=\{\left 
(\begin{array}{ll} a & b \\ c & d \end{array} \right) \in SL(2,\Z),\;\;\; \mbox{ with}\;\;\;\;
Nac, Nbd \;\;\;\;\;\mbox{ even}\;\;\;\}$.   Then:

(a) There exists a constant multiplier $m(g)$ such that the Toeplitz operator  
$U_g:= m(g) \Pi T_{\chi_g} \Pi$ is unitary.   The space of elements $H^2_{\Sigma}(N)$
of weight N relative to the center ${\cal Z}$ of $\Hb^{\red}_n$ may be identified with the
space $\Theta_N =\tilde{Th}^{i}_N$ of theta functions of degree N and the restriction $U_{g,N}:=
U_g |_{H^2_{\Sigma}(N)}$ defines the
 standard action
 (transformation law) of the element $g \in SL_{\theta}(2,\Z)$ on $\tilde{Th}^{i}_N$.

(b) The multipliers $m(g)$ may be chosen so that the
quantization maps $g \rightarrow U_{g,N}$ are projective representations
of $SL_{\theta}(2, \Z/N)$, and  indeed so that $U_{g,N}$ is the metaplectic
representation of $Mp_{\theta}(2,  \Z/N)$.

(c) The index of the symplectic map $g$ and contact transformation $\chi_g$
 in the sense of [Wei] equal zero.

 (d) If no eigenvalue of $g$ 
 is a root of unity, then the spectral data $\{e^{i\theta_{Nj}}, \phi_{Nj}\}$ of $U_{g,N}$ satisfy
the quantum mixing properties $({\cal MP }!)$ (cf. Definition 2a.6).

(e) One has the exact Egorov theorem: For $\sigma \in C^{\infty}(\R^{2n}/\Z^{2n}),$
 $U_g^* \Pi \sigma \Pi U_g = \Pi (\Pi_g \sigma \cdot \chi_g \Pi_g)\Pi$, where $\Pi_g$ is the
Toeplitz projector for the complex structure $g\cdot i.$  }
\medskip

The statements in (b)-(c) follow from that in (a).  The main point is that the Toeplitz
method produces the metaplectic representations.  As mentioned above, this is analogous
to the result of [D] which shows that the Toeplitz method produces the real metaplectic
representation.

In \S 6 we will present an exact trace formula for the traces of the quantized symplectic
torus automorphisms.  As noted in the introduction, the trace formula is classical
( [Kloo]), although the method of proof appears to be new.  
\medskip

\noindent{\bf Theorem E}{\it \;\;In the notation of Theorem D:
The multiplier $m(g)$ can be chosen so that the
 trace of the quantized cat map $U_{g,N}$ is given by
$$Tr U_{g,N} =  \frac{ 1}{\sqrt{det(I-g)}}
 \sum_{[(m,n)] \in \Z^{2n}/ (I-g)^{-1}\Z^{2n}}
 e^{i \pi N [\langle m,n \rangle - \sigma ((m,n), (I-g)^{-1} (m,n))]}$$
The square root is defined by the standard analytic continuation (cf. [F], \S 6).}
\medskip

This trace formula can be (and has been) used to analyse the fine structure of the
spectra of quantized cat maps.
The simplest case is that of  the elliptic element
$S=\left( \begin{array}{ll} 0 & 1 \\ -1 & 0 \end{array} \right)$.  Its quantization (on
theta functions of degree N) is equivalent to the finite Fourier transform
 $F(N)$ on $L^2(\Z/N)$ (cf. [A.T]. The trace formula then reads:
$$\frac{1}{\sqrt{N}} \sum_{r=0}^{N-1} e^{2 \pi i \frac{r^2}{N}} = \frac{1}{\sqrt{2}}
e^{i \pi /4} (1 + (-i)^N).$$
From this formula one can deduce that the eigenvalues of $F(N)$ are $\pm 1, \pm i$
with essentially equal multiplicities (loc.cit. ).  It follows that the `pair
correlation function' for the quantization $U_S$ of $S$ is a sum of delta functions.  

The above trace formula has also been studied previously in the physics 
literature, especially by Keating [Kea],
 to analyse the fine structure of the spectra of the $U_{g,N}$'s when $g$ is a
 hyperbolic automorphism.  In this case the eigenvalues become uniformly distributed
on the circle as $N\rightarrow \infty$.  On the scale of the mean level spacing, however,
the spectra of the $U_{g,N}$'s behave very 
erratically as $N\rightarrow \infty$:  For each N, there
exists a minimal positive integer $\tau(N)$, known as the quantum period, with the property that 
$U_{g,N}^{\tau(N)} = e^{i \phi(N)} Id.$  The eigenvalues $e^{i\theta_{Nj}}$ are therefore 
translates by $e^{i \phi(N)}$ of the   $\tau(N)$th roots of unity.   The erratic aspect is that
the period $\tau(N)$ depends on the factorization of N into primes and hence is very irregular
as a function of N.  Moreover the multiplicities
 $m_{Nj}$ of the  eigenvalues $e^{2 \pi i \frac{j}{\tau(N)} + \phi(N)}$ 
seem to be evenly distributed as $j$  varies over $\{0,1, \dots, \tau(N)-1\}$ [Kea].  It
follows that they tend to infinity at the erratic rate of $N / \tau(N).$  

The eigenvalue pair correlation problem for quantized hyperbolic cat maps thus involves some
intricate questions of number theory, while that for quantized elliptic maps is rather
trivial.  There are however some relatively interesting intermediate cases  whose
pair correlation functions can  be analyzed by means of the  trace formulae of the
above type (and we hope to do so in a future article).

\section{  Background }

\subsection*{ 2a: Review of quantum ergodicity and  mixing}

We begin by reviewing the notions of {\it quantized abelian system} and of {\it quantum
ergodic system} from [Z.1], and explain how they apply
in the present context.  We also review the mixing notions of [Z.2-3].

A quantum dynamical system is a $C^*$-dynamical system $({\cal A}, G, \alpha)$, where
${\cal A}$ is a unital, separable $C^*$-algebra and $\alpha: G \rightarrow Aut({\cal A})$
is a representation of G by automorphisms of ${\cal A}.$  We assume that ${\cal A}$ acts
effectively on a Hilbert space ${\cal H}$ and that there exists a unitary representation $U$ of
G such that $\alpha_g(A) = U_g^* A U_g.$  In other words, we assume the system is covariantly
represented on ${\cal H}.$

Since $G= S^1 \times \Z$ in this paper, we will assume G is an abelian; moreover we will
assume that the spectrum $\sigma(U)$ is discrete in the set Irred(G)(=$\Z \times S^1$ here) of
irreducible representations of G.  (In fact, in the Toeplitz examples the spectrum will not only be
discrete but will have the strong asymptotics properties described in Theorem A.)
We denote by ${\cal H}=\bigoplus_{\sigma \in \sigma(U)}
 {\cal H}_{\sigma}$ the isotypic decomposition of ${\cal H}$, by $\Pi_{\sigma}$
 the orthogonal projection onto ${\cal H}_{\sigma},$ and by $\omega_{\sigma}$ the invariant
state $\omega_{\sigma}(A)=\frac{1}{rk\Pi_{\sigma}} Tr \Pi_{\sigma}A.$   

 We then say:

\subsection*{2a.1 Definition}{\it  $({\cal A}, G, \alpha)$ is {\em  quantized abelian\/} if the
microcanonical ensembles 
$$\omega_E := \frac{1}{N(E)} \sum_{E(\sigma) \leq E} rk(\Pi_{\sigma})\omega_{\sigma}$$
have a unique weak-\* limit as $E\rightarrow \infty,$ and if the $C^*$-dynamical system
$(\pi_{\omega}({\cal A}), G, \alpha_{\omega})$ associated to $\omega$ by the GNS construction
is abelian.}

Here, $rk$ is short for ``rank", N(E)=$\sum_{|\sigma| \leq E} rk(\Pi_{\sigma})$, and the sum
runs over $\sigma \in \sigma(U)$ of energy $E(\sigma)$ less than E, with $E(\sigma)$ 
essentially the distance from $\sigma$ to the trivial  representation. 
 We
regard $\omega$ as the classical limit (state) of the system, or 
$({\cal A}, G, \alpha)$ as the quantization of the associated GNS system.  The relevant
notions simplify a good deal in the case of the Toeplitz systems of this paper, and will
be further discussed in \S2.b.  For general discussion, including generalities on classical
limits of Toeplitz systems, see [Z.1].

We also say:

\subsection*{(2a.2) Definition}{\it  A quantized abelian system $({\cal A}, G, \alpha)$ is 
{\em quantum ergodic \/} if there exists an operator K in the von-Neumann algebra closure of ${\cal A}$
such that
$$<A> = \omega(A) I + K \;\;\;\;\;\;\; \mbox{with} \;\;\;\;\omega_E(K^*K)\rightarrow 0.$$}
Here, $<A>$ is the time average of A, 
$$<A> = w-\lim_{T\rightarrow \infty} <A>_T$$
where 
$$<A>_T:=\int_G \psi_T(g) \alpha_g(A)dg,$$
with $\psi_T$ an ``M-net" (approximate mean) for G.  In the case
$G=S^1 \times \Z$, $\psi_T(g)dg = \frac{1}{T} \chi_{[-T,T]}(t)dt d\theta$ where $d\theta $ 
(resp. dt) is Lebesgue measure on $S^1$ (resp. counting measure on $\Z$).

 This notion of quantum ergodicity is equivalent to a condition on the eigenfunctions of
the quantum system.   To state it, we
recall that in 
 a generalized {\it quantized abelian} system the group $G$ is assumed to have the form
$T^n \times\ R^k \times \Z^m$ (with $T^n$ the n-torus).  Hence
 the irreducibles are 1-dimensional,  of the
form $\C \phi_{\chi}$ where $\phi_{\chi}$ is an
 eigenfunction corresponding to a character $\chi$ of $G$.  By our assumptions
above, the set of such characters is discrete in the dual group $\hat{G}$ and we 
enumerate them in a sequence $\chi_j$ according to their distance $E(\chi_j)$ to the
 trivial representation .  The corresponding
eigenfunctions will be denoted $\phi_j$.  To each is associated an ergodic invariant state
$\rho_j$ of the quantum system, namely the vector state $\rho_j(A)=(A \phi_j,\phi_j).$
The criterion above of quantum ergodicity is equivalent to the following:
$$\exists {\cal S} \subset spec(U): D^*({\cal S})=1\;\;\;\;\;\;\;\;w-\lim_{j \rightarrow \infty,
\chi_j \in {\cal S}} \rho_j = \omega.$$
Here $D^*({\cal S)}$ is the density of ${\cal S}$ (see [Z1]). 

We have:

\subsection*{ (2a.3) Theorem }([Z.1, Theorem 1-2]).  {\it Suppose $({\cal A}, G, \alpha)$ is quantized
abelian.  Then: if $\omega$ is an ergodic state, the system is quantum ergodic.}
\smallskip

There is a more refined result due  to Sunada [Su] and to the author [Z.1].  
\medskip

\subsection*{(2a.4) Theorem }([Su][Z.1, Theorem 3]) {\it With the same notation and assumptions
as in Theorem A.1:  ergodicity of $\omega$ is equivalent to quantum ergodicity of
$({\cal A},G,\alpha)$ together with the following strong ergodicity property:
$$\lim_{T\rightarrow \infty}\lim_{E\rightarrow \infty} \omega_E(\langle A \rangle_T^*A)
=\lim_{E\rightarrow \infty}\lim_{T\rightarrow \infty} \omega_E(\langle A \rangle_T^*A)
=|\omega(A)|^2.\leqno({\cal EP!})$$
Further, ${\cal EP!}$ is equivalent to:
 $$(\forall \epsilon)(\exists \delta)
\limsup_{E \rightarrow \infty} \frac{1}
{N(E)} \sum_{{j \not= k: E(\chi_j), E(\chi_k) \leq E}\atop
{ |E(\chi_j) - E(\chi_k)| < \delta}} |\rho_{jk}(A)|^2 < \epsilon \leqno({\cal EP!})$$
where
$\rho_{ij}(A) = (A \phi_i,\phi_j) = Tr A \cdot \phi_i \otimes
\phi_j$.}
\medskip

 We now recall
some analogous definitions and results in the case where the geodesic flow
is weak-mixing.  

Quantum weak mixing
 has to do with the mean Fourier transform
$$\hat{A} (\chi):= w-\lim_{T\rightarrow \infty} \hat{A}_T(\chi)$$
of observables $A \in {\cal A}$, where  $\chi \in\hat{G}$  
where $$\hat{A}_T(\chi)=
\int_G \psi_T(g)
 \alpha_g(A) \overline{\chi}(g) dg$$  is the partial mean Fourier transform,
 and where the limit is taken in the
weak operator topology.  (When the
expression for $A$ gets too complicated we often write  ${\cal F}_T(A)(\chi)$  for
this transform and similarly for the limit as $T\rightarrow \infty$).
  The following generalizes the
definition  of a quantum weak mixing system given in [Z.2] for the case of the systems
$(\Psi^o(M), \R, \alpha)$, with $\Psi^o(M)$ the $C^*$-algebra of bounded pseudodifferential
operators over a compact manifold $M$ and with  $\alpha_t(A) = U_t^* A U_t$ 
the automorphisms of conjugation by the
wave group $U_t:= e^{it\sqrt{\Delta}}$ of a metric $g$ on $M$:
\medskip

\subsection*{\bf (2a.5) Definition} {\it A quantized abelian system is {\em
 quantum weak mixing \/}
if, for  $\chi \not= 1$, 
 $$\limsup_{E\rightarrow \infty} \omega_E(\hat{A}(\chi) \hat{A}(\chi)^*) =0 .\leqno({\cal MP})$$}
\medskip

 As in the case of ergodicity there are also sharper weak mixing conditions which
involve the  
  partial mean Fourier transforms
 and the eigenfunctionals $\rho_{ij}(A):= ( A\phi_i,\phi_j)$ above.
We note that the eigenfunctionals of the automorphism group correspond to eigenvalues
in the ``difference spectrum" $\{\chi_j \overline{\chi}_k\}$.  For motivation and background
on quantum weak mixing we refer to [Z2,3] .
 In the following $\hat{G}^*$ denotes the set of non-trivial characters of $G$.
\medskip

\subsection*{(2a.6) Definition } {\it  A quantized abelian system has the {\em full weak mixing
property \/} if  in addition to ${\cal MP}$ it satisfies, for all $\chi \in \hat{G}^*$:

$$\lim_{T \rightarrow \infty}\lim_{E \rightarrow \infty} \frac{1}
{N(E)} \omega_E(\hat{A}_T(\chi)^*\hat{A}_T(\chi)) =0
\leqno({\cal MP!})\;\;\;\;\;\;\;$$}

We have:
\medskip

\subsection*{(2a.7) Theorem  ([Z.2,3]} {\it Let $({\cal A},G,\alpha)= (\Psi^o(M), \R,
\alpha)$ with $\alpha$ the automorphism of conjugation by the wave group $U_t$ of
a Riemannian metric $g$ on $M$. Then:  the geodesic flow $G^t$ is weak mixing
on the unit cosphere bundle $S^*M$ with respect to Liouville meassure $\mu$ if and
only if the
quantum system  $({\cal A},G,\alpha)$ has the mixing properties ${\cal MP}$
and  ${\cal MP!}$.}
\medskip

The same statement is true in the case of the Toeplitz systems of this paper,  and
with more or less the same proofs.  For the sake of brevity we will restrict our
attention to the following concrete consequence of weak mixing for the matrix
elements $\rho_{ij}(A)$  of observables:
\medskip

\subsection*{(2a.8) Corollary ([Z.2,3])} {With the same notations and assumptions
as in Theorem (2a.7), we have:
$$(\forall \epsilon)(\exists \delta)
\limsup_{E \rightarrow \infty} \frac{1}
{N(E)} \sum_{{j\not= k: E(\chi_j), E(\chi_k) \leq \E}\atop
{ |E(\chi_j \overline{\chi_k \cdot \chi})| < \delta}} |\rho_{ij}(A)|^2 < \epsilon
\leqno({\cal MP!^*})\;\;\;\;\;\;\;$$}
\medskip

This Theorem will be generalized in the form of Theorem C to Toeplitz systems.

\subsection*{2b: Periodic contact manifolds and Toeplitz algebras}  We now introduce
the {\it quantized abelian} systems which play the principal role in this article: the ones
generated by periodic contact flows and quantized contact transformations.   The proof
that the quantizations can be unitarized will be postponed to \S3, where a good deal of further
background on Toeplitz operators and their symbols will be reviewed. 

The setting opens with a  compact contact manifold $(X,\alpha )$ .  The characteristic
distribution $ker d\alpha$ of $\alpha $ is one dimensional, and hence there
exists a vector field $\Xi$ on $X$ such that $\alpha (\Xi) = 1$ and
$\Xi \bot d\alpha  = 0$.  We will make the 
\medskip

\subsection*{(2b.1)  Assumption 1}  The characteristic flow
$\phi^t$ of $\Xi$ is periodic.  
\medskip

This assumption is satisfied in the motivating examples from geometric quantization
theory and complex analysis.
  Thus, suppose $L\rightarrow M$ is a
holomorphic line bundle over a compact complex manifold, and let
$\|\cdot\|$ be a hermitian metric on $L$. Then the disc bundle
$\Omega = \{(x,v):|v|_x<1\}$ is a strictly pseudo convex domain,
whose boundary $\partial \Omega$ has a natural contact structure
with periodic characteristic flow.  Of particular interest here are the line bundles
which arise as ``pre-quantum line bundles" over Kahler manifolds in Kahler quantization.
 See [A.dP.W][B.G][We] for examples and further discussion.

Now let
$$\Sigma:=\{(x, r\alpha _x):r>0\}\subset T^*X\backslash
0\;.\leqno(2b.2)$$
Then $\Sigma $ is a symplectic cone , and according to Boutet-de-Monvel and Guillemin [B.G]
 always has a Toeplitz structure
$\Pi_\Sigma $, that is, an orthogonal
projection with wave front along the graph of the identity on $\Sigma$, 
and with the microlocal properties of the Szego projector onto boundary values
of holomorphic functions on a strictly pseudoconvex domain. 

 The algebra of
concern is then the Toeplitz algebra $\T^\circ_\Sigma $
 associated to $\pis $.  By definition, this is the algebra of operators $\pis A \pis$ on
$L^2(X)$ with $A\in \Psi^o(X)$ (i.e. the algebra of zeroth order
 pseudodifferential operators over X).
The range of $\pis$ will be denoted $H^2_{\Sigma}$, i.e.
$$\Pi_\Sigma:L^2(X) \rightarrow H^2 _\Sigma  (X).$$ 
It is  clear that the 
Toeplitz algebra is effectively represented on this Hilbert space.

As mentioned above, the group $G$ of concern will be $S^1\times \Z$.  The circle $S^1$ will operate
on $L^2 (X, d\nu)$ by:  $W_t \cdot f(x) = f(\phi^{-t} x)$.  Here $d\nu$
is the normalized volume form determined by $\alpha $, i.e.\
$d\nu= c\alpha \wedge(d\alpha )^{n-1}$ for some $c>0$, where
$\dim X = 2n+1$.  We may (and will) assume that $\pis $ is
chosen so that $[\pis, W_t]=0$ ([B.G, Appendix]).  Then $S^1$ will operate on
 $ H^2_{\Sigma}$.  Its generator $\frac{1}{i} D_{\Xi}$ compresses to the Toeplitz
operator
$\Pi_{\Sigma} \frac{1}{i}D_{\Xi} \Pi_{\Sigma}$.  The symbol of this Toeplitz operator
is the function $\Phi : \Sigma \rightarrow \R, \Phi(x, \xi) = \langle \xi,
\Xi\rangle.$

The group $\Z$ will act by powers of a quantized contact
transformation.  By definition this will be a unitary operator
$U_{\chi}:H^2_\Sigma (X) \rightarrow H^2_\Sigma  (X)$ of the form:
$$U_\chi = \Pi_\Sigma  T_\chi A\pis\leqno(2b.3)$$
where $T_\chi f(x) = f(\chi^{-1}(x))$ and where $A \in
\Psi^\circ_{\pis}$.  We will make the 
\medskip

\noindent{\bf (2b.4)  Assumption 2.  $\Phi > 0$ and $[T_\chi, W_t]=0$ }
\medskip

These assumptions are also satisfied in the examples from Kahler quantization theory.
They imply that $\chi$ descends to
a symplectic automorphism $\chi_{\cal O}$ of the quotient ${\cal O}= X/S^1$ of $X$ by
the action of $\phi^t.$  They also imply that $\Pi_{\Sigma} \frac{1}{i} D_{\Xi} \Pi_{\Sigma}$
is an elliptic Toeplitz operator, and hence that the isotypic subspaces $H^2_{\Sigma}(N)$
are finite dimensional.  Vice-versa they hold if $(X, \alpha)$ is the prequantum $S^1$-
bundle of an integral symplectic manifold [B.G, Lemma 14.9].

To show that (2b.3) is non-vacuous we will prove the: 

\subsection*{(2b.5) Unitarization Lemma}  {\it Let $\chi$ be a contact
transformation of a contact manifold $(X,\alpha )$ satisfying the Assumptions 1-2.
  Then there exists a symbol $\sigma_A \in C^{\infty}({\cal O})$ determined in a canonical
way from $\chi, \Pi_{\Sigma}$
 and a canonically constructed
operator $A\in\Psi^\circ_{\pis}$ with principal symbol $\sigma_A$ such
that $[A, \frac{1}{i} D_\Xi]=0$ and such that $U_{\chi,}$ in (2b.3) is unitary
on $H^2_{\Sigma} $ (at least on the complement of a finite dimensional
subspace).}
\medskip

Granted the Unitarization Lemma, the $C^*$-dynamical system of concern 
will be $({\cal T}^o_{\Sigma}, S^1\times\Z,
\alpha)$ where $\alpha_{k,t}$ is conjugation by $U_{\chi}^k W_t.$  By the composition
theorem of [B.G], such conjugations are automorphisms of the Toeplitz algebra.

The principal symbol of $\pis A \pis$ may be idenfitied with $\sigma_A|_{\Sigma};$ a
more complete description of the symbol will follow in the next section.  The symbol
algebra of ${\cal T}_{\Sigma}^o$ (zeroth order Toeplitz operators) may then be identified
with smooth homogenous functions of degree 0 on $\Sigma;$ hence with functions on X.
In the $C^*$-closure one  gets all the continuous functions.  

Many of the notions involved in the definition of quantized abelian systems in (${\bf \S2.a}$)
 simplify a good deal for these Toeplitz systems.  First,
the irreducibles correspond to characters $\chi_{(N,\tau)}:=
e^{2\pi iNt}\otimes e^{ik\tau}$ of $S^1 \times \Z$ and have the form $\C \phi_{(N,i)}$ where
$U_{\chi}^k W_t.\phi_{(N,j)}=e^{2\pi iNt} e^{ik \theta_{(N,j)}}\phi_{(N,j)}$.  The energy 
$E(\chi_{(N,\tau)})$ is defined to be $N$.  Hence the microcanonical ensembles
$\omega_E$ have the form
$$\omega_E = \frac{1}{N(E)} \sum_{N=1}^E d_N \omega_N$$
where  $E \in \N$ and where $\omega_N$ is the degree n ensemble defined by
 $$\omega_N := \frac{1}{d_N}
\sum_{i=1}^{d_N} \rho_{(N,i)}\;\;\;\;\;\;\;\;
\mbox{with}\;\;\;\;\; \rho_{(N,i)}(A):= (A\phi_{(N,i)}, \phi_{(N,i)}).$$  The ensembles
$\omega_E$ are equivalent to the ensembles
$$\tilde{\omega}_E:= \frac{1}{E} \sum_{N=1}^E \omega_N$$
in the sense that $\omega_E(A) = \tilde{\omega}_E (a) + o(1)$ as $E\rightarrow \infty$.
This follows easily from the fact that $d_N \sim N^{dimX -1}$ has polynomial growth
(for similar assertions see [Z1,4].)  In fact, the order n ensembles $\omega_N$ have
sufficiently well-behaved asymptotics that we will not need to further average over
all $N$.  This accounts for the stronger kinds of results available in the Toeplitz case.

\subsection*{(2b.6)Proposition}{\it $({\cal T}^o_{\Sigma}, S^1 \times \Z, \alpha)$ is a quantized
abelian system, with classical limit system $(C(X), S^1 \times Z, \alpha_{\omega})$, 
where $\alpha_{\omega (t, k))}$ is  conjugation  by  $T_{\chi}^k \cdot  W_t.$}

\subsection*{Proof}

With the assumptions (1)-(2) above, as well as the temporary assumption of
the Unitarization Lemma ,
the isotypic decomposition of $H^2_{\Sigma}$ is just its decomposition into joint
eigenspaces for $(U_{\chi}, W_t)$,  and the  weight spaces $H^2_{\Sigma}(N)$
 are just the eigenspaces of
$W_t$ corresponding to the characters $e^{2\pi i N t}$.  Let $\Pi_N$ denote
 the associated orthogonal
projection, and let $d_N = dim H^2_{\Sigma}(N) = rk(\Pi_N).$  Then we have (with $E
\in {\bf N}), A \in {\cal T}^o_{\Sigma},$
$$\omega_N(A) =\frac{1}{d_N} Tr \Pi_N A$$
where of course
$$ \omega_E (A) = \frac{1}{N(E)} \sum_{ 0\leq  N\leq E} Tr A \Pi_N$$
and of course $N(E) = \sum_{N \leq E} d_N.$  The asymptotics of $\omega_N(A)$ follow in a
standard way from the singularity asymptotics at $t=0$ of the dual sum
$$\sum_{N\geq 0} (TrA \Pi_N)e^{2\pi i N t} = \sum_{N\geq 0} d_n \omega_N(A) 
e^{2\pi i Nt} = Tr \pis A W_t. \leqno (2b.7)$$
The composition theorem for Fourier Integral operators and Hermite Fourier Integral 
operators [B.G., \S7] shows that the trace is a Lagrangean distribution with
 singularity only at $t=0$ and
with principal symbol at the singularity  given by
$$\omega(A) = \int_{X} \sigma_A d\nu.$$
Here, $d\nu = \alpha \wedge (d\alpha)^{n-1},$ and as above $\sigma_A$ is
identified with a scalar function on $X$.  In fact the trace (2b.7) is
a Hardy distribution on $S^1$ so one can conclude, simply by
 comparing Fourier series expansions,
 that $$\omega_N(A) \sim \omega(A) + O(N^{-1})\leqno(2b.8)$$ for smooth
Toeplitz operators $A$ (see [B.G, \S 13] for details of this argument).   It 
obviously follows that $\omega_E(A), \tilde{\omega}_E(A) \rightarrow \omega(A).$
Since (2b.8) is much stronger we will henceforth use it as the key property of
the Toeplitz system. 

To complete the proof, we only need to identify the classical limit system precisely.
From the composition theorem we have
$$\sigma(\alpha_{t,k}(A))=\sigma_A \cdot(\phi^t \cdot \chi^k).\leqno(2b.9)$$
and hence  need only to identify the GNS representation with the symbol map. However,
it is clear that for smooth elements $A \in {\cal T}_{\Sigma}^o$ (i.e. not in the norm-closure),
$\omega(A^*A)=0$ if and only if $\sigma_A = 0$, hence the ideal ${\cal N}=\{A:\omega(A^*A)=
0\}$ is the norm closure of ${\cal T}^{-1}_{\Sigma}$, namely the ideal ${\cal K}$ 
of compact operators in the algebra.  However one has the exact sequence
$$0\rightarrow {\cal K} \rightarrow {\cal T}^o_{\Sigma} \rightarrow C(X) \rightarrow 0$$
where the last map is the symbol map [D]. Hence ${\cal T}^o_{\Sigma}/{\cal N}$, closed 
under the inner product induced by $\omega$ is precisely $L^2(X, d\nu),$ and the induced
automorphisms are those of (2b.9). \qed

\section{ Symplectic spinors and proof of the Unitarization Lemma }

The main point of this Lemma is to determine the principal symbol $\sigma_A$ of the
pseudodifferential operator $A$ in (2b.3) which unitarizes the Toeplitz translation
by $T_{\chi}$.   The rest follows by use of the functional calculus.  The length of the
proof is due mainly to the review it contains of symplectic spinors
and symbols of Toeplitz operators.  We will see this symbol reappearing in the course
of the proof of Theorem D.  

\begin{pf}  By [B.G, Theorem 7.5] any operator of the form
(2b.3) is a Fourier Integral operator of Hermite type with wave
front along the graph of $\chi|_\Sigma $.  Here, $\chi$ is understood
to extend to $\Sigma $ as a homogeneous map of order $1$.
Therefore, the main point is to construct $A$ so that $U_{\chi}$ is unitary,
and so that it commutes with the other operators. Consider first the unitarity.  At the
principal symbol level, this requires
$$\sigma(\pis  A^* T^{-1}_\chi \Pi_\Sigma  T_\chi A\Pi) =
\sigma (\Pi_\Sigma )\;.\leqno(3.1)$$

To solve  (3.1) , we will have to go further into the
symbol algebra of $\T^\cdot_\Sigma $:  We first recall that the principal
symbol of a Hermite operator is a ``symplectic spinor" on $\Sigma $.  In
other words, a homogeneous section of the bundle
$$\Spin(\Sigma ^\#)\simeq \Lambda^{\half} (\Sigma ^\#)\otimes {\cal
S}(\Sigma ^{\#\bot} /\Sigma ^\#)$$
where ([B.G, p.41][G.2])

\subsection*{3.2}  (i)  $\Sigma^\# = \{ x, \xi, x, -\xi):(x,\xi)\in
\Sigma\}$

(ii)  $\Lambda^{\half}$ is the $\half$ form bundle

(iii)  $\Sigma^{\#\bot} /\Sigma^\#$ is the symplectic normal bundle of
$\Sigma^\#$

(iv)  ${\cal S} (\Sigma^{\#\bot}/\Sigma^\#)$ is the bundle of Schwartz
vectors along $\Sigma^{\#\bot}/\Sigma^\#$.
\medskip

In the case at hand, $\Lambda^{\half} (\Sigma^\#)$ has a natural
trivialization coming from the symplectic volume $\half$-form on
$X$.  Hence we can ignore it.  Also,
$(\Sigma^{\#\bot}/\Sigma^\#)_{(p,-p)}$ is the sum $(\Sigma_p)^\bot
\oplus (\Sigma^\bot_p)$ where $(\Sigma_p)^\bot$ is the symplectic
orthogonal complement of $\Sigma_p :=T_p\Sigma$ in $T_p(T^*X)$.  For
each choice of symplectic basis of $(T_p\Sigma)^\bot$, one has
identifications
\begin{gather*}
(T_p\Sigma)^\bot \simeq \R^\ell \otimes (\R^\ell)^*\tag{3.3}\\
{\cal S} (\Sigma^{\#\bot} /\Sigma^\#)_{(p,-p)}\simeq {\cal S} (\R^\ell
\oplus \R^\ell)\;.\end{gather*}
Here, $\R^\ell \oplus \R^\ell$ is the Lagrangean subspace of
$(T_p\Sigma)^\bot \oplus (T_p \Sigma)^\bot$ indicated in (3.3) and
${\cal S}$  is the usual space of Schwartz functions.  A symplectic
spinor $\sigma $ can be identified in this way with the kernel
$\kappa_\sigma (p,\circ,\circ)$ of a smoothing operator
$T(\sigma ,p)$ on the Hilbert space $L^2(\R^\ell)$.

Consider in particular the Szeg\"o--Toeplitz projector
$\Pi_\Sigma$.  According to [B.G, Theorem 11.2] its symbol
$\sigma (\Pi_\Sigma)$ may be described as follows:  First,
$\Pi_\Sigma$ determines a homogeneous positive definite
Lagrangean sub-bundle $\Lambda$ of $\Sigma^\bot \otimes \C$ (the
complexified normal bundle of $\Sigma$ in $T^* (X)$).  For each $x\in
\Sigma$, $\Lambda_x$ then determines a unique (up to multiples)
vector $e_{\Lambda_{x}}\in {\cal S} (\Sigma^\bot_x)$, called the
vacuum vector corresponding to $e_{\Lambda_{x}}$.  Then $T(\sigma
(\Pi_\Sigma),x) = e_{\Lambda_{x}}\otimes e_{\Lambda_{x}}^*$, i.e.\
$\sigma  (\Pi_\Sigma)$ is the rank one projection onto $\C
e_{\Lambda_{x}}$.

To bring this somewhat down to earth, we note that $\Pi_\Sigma$ is
annihilated by an involutive system of $d = \half \dim \Sigma-1$
equations
\begin{gather*}
D_j\Pi_\Sigma \sim 0\quad(\mbox{modulo } \Psi^{-\infty})\tag{3.4}\\
[D_j, D_k]\sim \Sigma A^m_{jk} D_m\quad(A^m_{jk} \in
\Psi^\circ)\end{gather*}
similar to the tangential Cauchy--Riemann equations for the Szego
projector of a strictly pseudo convex domain.  Above,
$\Psi^{-\infty}$ is the algebra of smoothing operators on
$X$.  The characteristic variety of this system is $\Sigma$, and the
matrix $\frac{1}{i} \{\sigma (D_j),\sigma (D_k)\}$ is Hermitian
positive (or negative) definite along $\Sigma$.  Let $H_{\sigma_j}$ be the
Hamilton vector field of $\sigma_j:=\sigma (D_j)$ and set
$$\Lambda_x = \spann _{\C}\{H_{\sigma_j}:j=1,\ldots, d
\}\;.\leqno(3.5)$$
One can check that $\Lambda_x \subset \Sigma^\bot_x \otimes \C$,
that $\dim_{\C} \Lambda_x = \half \dim_{\C} \Sigma^\bot \otimes\C$
and that $\Lambda_x$ is involutive.  Hence, $\Lambda_x$ is a
Lagrangean subspace of $\Sigma^\bot_x\otimes \C$.

Now let $Mp(\Sigma^\bot)$ be the metaplectic frame bundle of
$\Sigma^\bot$: i.e.\ the double cover of the symplectic frame bundle
of $\Sigma^\bot$ corresponding to the cover $Mp (2n,\R)
\rightarrow Sp((2n,\R)$.  Then
$${\cal S} (\Sigma^\bot) = Mp (\Sigma^\bot ) \times_\mu {\cal S} (\R^\ell)$$
where $\mu$ is the metaplectic representation.  From this one can
transfer the Schr\"odinger representation $\rho$ of the Heisenberg
group on ${\cal S} (\R^\ell)$ to ${\cal S} (\Sigma^\bot)_x$ for each
$x$, and each metaplectic frame of $\Sigma^\bot_x$.  If $d\rho_x$
represents the derived representation at $x$, then one sees that
the symbol equations corresponding to (3.4) are
$$d\rho_x(H_{\sigma_j}) \sigma (\Pi_\Sigma) = 0\quad (x\in \Sigma
,\Xi_j\in \Lambda_x)\;.\leqno(3.6)$$
The vacuum state $e_{\Lambda_{x}}$ is the unique solution of the
similar system of equations on ${\cal S}(\Sigma^\bot)_x$.  Since
$\sigma (\pis)$ is a projector it must be $e_{\Lambda_{x}}
\otimes e_{\Lambda_{x}}^*$.

Next, return to (3.1).  By the composition theorem [B.G, 7.5], 
$$\sigma (\pis A^* T^{-1}_\chi \pis T_\chi A\Pi) = |\sa|^2\cdot
\sigma _{\Pi_{\Sigma}^{\circ}} \sigma (T^{-1}_\chi \pis
T_\chi)\circ \sigma _{\pis}\;.
\leqno(3.7)$$
Now $T^{-1}_\chi \pis T_\chi$ is also a Toeplitz structure on
$\Sigma$, since $\chi$ is a symplectic diffeomorphism of $\Sigma$. 
Hence $\sigma (T^{-1}_\chi \pis T_\chi)$ will be a rank one
projection $e_{\Lambda_{\chi}}\otimes e_{\Lambda_{\chi}}$ for
some Lagrangean sub-bundle $\Lambda_\chi$ of $\Sigma^\bot
\otimes \C$.  In fact
$$\Lambda_\chi = d\tilde\chi (\Lambda )\leqno(3.8)$$
where $\tilde \chi:T^*(X) \rightarrow T^* (X)$ is the natural lift
$(d\chi^t)^{-1}$ of $\chi$ to $T^*(X)$.  Note that $\tilde \chi
|_\Sigma =\chi|_\Sigma$, and since $\tilde \chi$ is symplectic,
\begin{gather*}d\tilde\chi:T\Sigma\otimes\C \rightarrow T\Sigma
\otimes \C\;,\\
d\tilde\chi:\Sigma^\bot \otimes \C\rightarrow \Sigma^\bot \otimes
\C\;.\end{gather*}
Also, $d\tilde\chi(\Lambda )$ is lagrangean sub-bundle of
$\Sigma^\bot \otimes \C$.  By the symbolic calculus, $\sigma
(T^{-1}_\chi \pis T_\chi)$ will have to solve (3.6) with $\Xi_j$
replaced by $d\tilde\chi(\Xi_j)$; hence (3.8).

Carrying out the composition of projections in (3.7), we conclude
that 
$$\sigma (\pis A^*T_\chi^{-1} \pis T_\chi A\pis ) = |\sa|^2\langle
e_{\Lambda _{\chi}},e_\Lambda \rangle|^2 e_\Lambda \otimes
e^*_\Lambda\;. \leqno(3.9)$$
To satisfy (3.1) it is sufficient to set:
$$\sa(x) = \langle e_{\Lambda _{\chi}},e_\Lambda
\rangle^{-1}\;.\leqno(3.10)$$ 
Of course, we must show that $\langle e_{\Lambda_{x}},
e_{\Lambda}\rangle (x)$ never vanishes.  In the model case $\R^\ell$,
$e_{\Lambda _{\chi}}$ and $e_\Lambda $ correspond to a pair of Gaussians
$\gamma_{Z_{1}}$ and $\gamma_{Z_{2}}$, where $\gamma_Z = e^{i\langle
ZX,X\rangle}$ for a complex symmetric matrix $Z = X +iY$ with $Y\gg 0$. 
It is obvious that $\langle \gamma_{Z_{1}},\gamma_{Z_{2}}\rangle$
never vanishes, since the Fourier transform of a Gaussian is never zero.

Now let $S^1$ act via $W_t$.  Since $\pis$ commutes with the
action, one may assume the operators $D_j$ in (3.4) commute with
the action (otherwise, one can average them).  Hence, the
Lagrangean sub-bundle $\Lambda $ is $S^1$-invariant, and since
$\chi$ commutes with the $S^1$ action, $\Lambda _\chi$ is also
$S^1$-invariant.  It follows from uniqueness of the vacuum vectors
(up to multiples) that $e_\Lambda $ and $e_{\Lambda _{\chi}}$ are
eigenvectors of the $S^1$ action.  Since they correspond under
$\tilde \chi$, they must transform by the same character.  It
follows that $\langle e_\Lambda , e_{\Lambda _{\chi}}\rangle$ is
$S^1$ invariant.  Hence, we have:
\subsection*{(3.11)} {\it $\sa$ is $S^1$-invariant.}
\smallskip

Now extend $\sa$ (in any smooth way) as a homogeneous function
of degree $0$ on $T^*X\backslash 0$, and let $A_1$ be any operator
in $\Psi^\circ_\Pi$ with symbol $\sa$.  By operator averaging
against $W_t$, we may assume $[A, \frac{1}{i} D_\Xi]=0$.  At this point,
$A_1$ satisfies:
\begin{gather*}
\begin{align*}[A_1, \pis ]&=0\\
[A_1, \frac{1}{i} D_\Xi]&=0\tag{3.12}\end{align*}\\[6pt]
U_1:=\pis T_\chi A_1\pis \quad\text{is unitary modulo}\quad{\cal
T}^{-1}_\Sigma\;.\end{gather*}
We now employ a simple argument of Weinstein [Wei] to correct
$A_1$ to define an operator $U_{(\chi,a)}$ which  is unitary.  In the following we
will pretend that the index $\ind(U_1)=0$.  If it is not, one has to work
on the orthogonal complement of a finite dimensional subspace.  This index is
an invariant of the contact transformation $\chi$ and hence is called the 
{\it index of $\chi$}. We will discuss it further in \S5.

By (3.12), $U^*_1U_1$ and $U_1 U^*_1$ are elliptic Toeplitz
operators, with principal symbols $\sigma (\pis)$.  Hence their
kernels are finite dimensional.  Let $K$ be an $S^1$-invariant
isometric operator from $\ker U_1\rightarrow \ker U^*_1$; let
$P$ denote the orthogonal projection onto $\ker U_1$.  Then $KP$
is a finite rank operator and
$$B_1 = U_1 +KP$$
is an injective Fourier Integral operator of Hermite type.  It
follows that $B^*_1 B_1$ is a positive Toeplitz operator with
symbol $\sigma (\pis)$.  Just as for pseudo differential operators,
there is a functional calculus for ${\cal T}^\circ_\Sigma$.  We may
express
$$B^*_1B_1 =\pis C\pis\quad\quad C\in \Psi^\circ _\Sigma$$
and then define
$$G = (B^*_1B_1)^{- \half} = \pis C^{-\half} \pis \in {\cal
T}^\circ_\Sigma\;.$$
Then set $U_{\chi} =B_1 G$.  It is unitary and satisfies all the conditions of the
lemma. \end{pf}

\section{Quantum ergodicity and mixing: Proofs of  Theorems A, B,  C}

We begin with the proof of the spectral dichotomy:

\noindent{\bf Proof of Theorem A}
\medskip

\noindent{\bf Proof of (i)}
The precise statement of (i) is that 
the eigenvalues $\{e^{2\pi i\theta _{Nj}} :j=1,\ldots, d_N\}$ with $d_N= dim H^2_{\Sigma}(N)$
become uniformly distributed on $S^1$ as $n\rightarrow \infty$ in
the sense that 
$$w- \lim_{N\rightarrow
\infty}\frac{1}{d_N}\Sigma^{d_N}_{j=1} \delta
(e^{2\pi i\theta_{Nj}})= d\theta\;\leqno(4.1)$$
or equivalently for $f\in C(S^1)$
$$\lim_{N\rightarrow \infty} \frac{1}{d_N}\sum^{
 d_N}_{j=1} f(e ^{2 \pi i\theta _{Nj}}) =  \int^{1}_0
f(e^{2\pi i\theta})d\theta\;.\leqno(4.2)$$
Of course, it suffices to let $f(z) = z^k (k\in \Z)$, and to prove that
the left side tends to $0$ if $k\neq 0$.  But if $f = z^k$, the left
side is 
$$\lim_{N\rightarrow \infty}\frac{1}{d_N}\Tr(U_{\chi,N}^k)
\Pi_N \leqno(4.3)$$
where as above $\Pi_N:H^2_{\Sigma}(X) \rightarrow H^2_{\Sigma}(N)$ is the orthogonal
projection.  The limit can be obtained from the singularity at
$\theta=0$ of the trace
$$\Tr W_\theta U_{\chi}^k = \sum^\infty_{N=0} e^{2\pi i N\theta}
\Tr(U_{\chi,N }^k) \Pi_N\;.$$
By the composition calculus of Fourier Integral and Hermite
operators [B.G], the singularities of the trace occur at values
of $\theta$ for which $e^{2\pi i\theta} \cdot \chi^k$ has non-empty
fixed point set.  It is clear that $\theta$ must equal zero, and that
the fixed point sets consists of the fibers over the fixed points of
$\chi^k$ on ${\cal O}$.  This is a finite subset if $k\neq 0$, and
hence the singularity is of the type $(t + i0)^{-1}$ ( compare
[B.G, Theorem 12.9]).  It follows that $\Tr(U_{ \chi, N} )^k$ is
bounded as $N\rightarrow \infty$ if $k\neq 0$ (compare [BG, Proposition 13.10]).  On
the other hand, 
$d_N=\dim H^2_{\Sigma}(N)\sim N^{(\dim X+1)/2-1} $ [loc.\ cit.], so the limit
(4.3) is zero unless $k=0$. \qed
\medskip

\noindent{\bf Proof of (ii)}:  If $\chi$ is periodic, then the whole group $G$ generated
by the contact flow and by $\chi$ is compact, and as mentioned above
the Toeplitz structure $\Pi_{\Sigma}$ may be constructed
to be invariant under it.  Hence the unitary quantization of $\chi$ is simply
$$U_{\chi}:= \Pi_{\Sigma} T_{\chi} \Pi_{\Sigma}$$
and $U_{\chi}^k = \Pi_{\Sigma} T_{\chi^k} \Pi_{\Sigma}.$  It follows that $U_{\chi}^p =
\Pi_{\Sigma} $
(the identity operator on $H^2_{\Sigma}$) and hence its eigenvalues are pth roots
of unity. \qed
\medskip

We now turn to the quantum ergodicity and mixing theorems.  The ergodicity theorems
follow almost immediately from the results of [Z1].

\subsection*{ Proof of Theorem B}

\subsection*{Proof}  By Proposition (2b.6),
 $({\cal T}^o_{\Sigma}, S^1 \times \Z, (W_t, U_{\chi}))$ is a quantized
abelian system and by assumption the classical limit system is ergodic.  
Except for one gap, the statement then
follows from Theorems 1-3 of [Z1].

The gap is that we are using the more localized ensembles $\omega_n$ rather than the
microcanonical ensembles $\omega_E$.  However, the only properties of $\omega_E$ used 
in [Z1] are that they form a sequence of  invariant states satisfying
 $\omega_E \rightarrow \omega$.  Since this was also proved for the degree n ensembles
$\omega_n$
 in Proposition (2b.6) (see
(2b.8), the proof of Theorem B is complete.
 \qed

\subsection*{Remark} {The ergodicity assumption is equivalent 
 to the ergodicity of $\chi$ on $({\cal O}, \mu)$ }

The following theorem states that if $\chi$ is weak mixing on $({\cal O},\mu)$, then
the quantum system has the full weak mixing property of Definition (2a.6) in the
even stronger form involving the degree n ensembles.  There is a notational overlap
in that we are writing $\chi$ both for characters and for the contact map; both
are conventional and we do not believe this should cause any confusion.

\subsection*{Proof of Theorem C}

\subsection*{Proof}  First, the weak mixing property of $\chi$ on $({\cal O}, \mu)$ is
equivalent to the statement that 
$$\lim_{M \rightarrow \infty} || {\cal F}_M (\tau) f||_{L^2} = 0\;\;\;\;\;(\forall f \bot 1)
\leqno(4.4)$$
where $${\cal F}_M(\tau) : L^2({\cal O}, d\mu) \rightarrow L^2({\cal O}, d\mu)$$
is the partial mean Fourier transform
$${\cal F}_M(\tau) f = \frac{1}{2M} \sum_{-M}^M e^{-im \tau} T_{\chi}^m f.$$
Indeed,  we have
$$\lim_{M \rightarrow \infty} ||{\cal F}_M(\tau) f - P_{\tau}f||_{L^2} =0$$
for all $f \in L^2$, where $P_{\tau}$ is the orthogonal projection onto the eigenspace of
$T_{\chi}$ of eigenvalue $P_{\tau}$.   On the other hand if $\chi$ is weak mixing, then
$P_{\tau} f = 0$ for all $f\bot 1$ since the unitary operator
$$T_{\chi} : L^2({\cal O}, d\mu) \rightarrow L^2({\cal O}, d\mu)$$
$$T_{\chi} f( o) = f(\chi^{-1} (o)$$
has no $L^2$-eigenfunctions other than constants. Henceforth we only consider non-trivial
characters ($\tau \not= 0$) since the case $\tau = 0$ is covered in the ergodicity theorem
above.

One connection to the quantum theory is thru the partial mean Fourier transforms
$$\hat{A}_M(\chi) = {\cal F}_M(\chi) A:= \int_G \psi_M(g) \alpha_g(A) \overline{\chi}(g) dg$$
of observables $ A \in {\cal T}^o_{\Sigma}$.  To simplify, we recall that
 without loss of generality,  an element
of  $ {\cal T}^o_{\Sigma}$ may be assumed to be of the form
 $\Pi_{\Sigma} A \Pi_{\Sigma}$ with
$A \in \Psi^o (X)$, with $[A, W_t]=0, [A, \Pi_{\Sigma}] \sim 0$ [B.G].  As above, we also
 write  characters $\chi$ in
the form $e^{2\pi i N t}\otimes e^{ik \tau}$ with $e^{2\pi iN t} \in S^1.$  We further note that
the quantum mixing condition stated in the Theorem 
 concerns only the diagonal blocks $\Pi_N A \Pi_N$ whose partial mean Fourier 
transforms have the form
$${\cal F}_M (\chi) \Pi_N A \Pi_N =  \frac{1}{2M} \sum_{m= -M}^M e^{-im \tau}\int_{S^1}
e^{-2\pi iNt} W_t^* U_{\chi}^{-m}
\Pi_N A \Pi_N U_{\chi}^m W_t dt.$$
Since $[U_{\chi}, \Pi_N]=[W_t,\Pi_N]=0, W_t\Pi_N = e^{2\pi iN t}\Pi_N$, 
the conjugates $\alpha_g (\Pi_N A \Pi_N)$ are constant in $t$ and hence
  ${\cal F}_M(\chi)\Pi_N A \Pi_N=0$ 
unless the character $\chi$ has the form $1 \otimes e^{ik\tau}$.  In the latter
case, the partial mean Fourier transforms of the blocks simplifies to
$${\cal F}_M (\tau) \Pi_N A \Pi_N =  \frac{1}{2M} \sum_{m= -M}^M e^{-im \tau} U_{\chi}^{-m}
\Pi_N A \Pi_N U_{\chi}^m\leqno(4.5)$$
which begin to look very much like their classical counterparts.  The resemblence is
of course made even closer by use of the Egorov  theorem for Toeplitz operators [B.G],
which implies that $U_{\chi}^{-m} \Pi A \Pi U_{\chi}^m \in {\cal T}^o_{\Sigma}$ with
principal symbol equal to $ T_{\chi}^m\sigma_{\Pi A \Pi} .$ 

We now make the key observation:
$$\lim_{N\rightarrow \infty} \frac{1}{d_N}
 ||{\cal F}_M(\tau) \Pi_N A\Pi_N||_{HS}^2 = ||{\cal F}_M(\tau)\sigma_A||^2_{L^2}\leqno(4.6)$$
where $\sigma_A$ denotes the function on ${\cal O}$ induced by the principal symbol
of $\Pi A \Pi.$  This follows from the Egorov theorem combined with a special case
of the  Szego limit theorem for Toeplitz operators:
$$\lim_{N\rightarrow \infty} \frac{1}{d_N}||\Pi_N A \Pi_N||^2
= \frac{1}{\mu({\cal O})} \int_{{\cal O}} |\sigma_A|^2 d\mu \leqno(4.7)$$
which holds if $\sigma_A|_{\Sigma}$ is invariant under the contact flow (see [B.G,
Theorems 6 and 13.11].)  

It follows from (4.4) and (4.6) that if $\chi$ is weak mixing and $\tau\not= 0$, then
$$\lim_{M\rightarrow \infty}\lim_{N\rightarrow \infty} \frac{1}{d_N}
 ||{\cal F}_M(\tau) \Pi_N A\Pi_N||_{HS}^2 =0\leqno(4.8)$$
 for all smooth  $A$  in the Toeplitz algebra.
Let us now express (4.8) in terms of the eigenfunctions $\phi_{(N,i)}$ of $U_g$ and
in terms of the eigenfunctionals $\rho_{(N,ij)}(A):= (A \phi_{(N,i)}, \phi_{(N,j)})$
of the automorphisms $\alpha_g.$  We have
$$\rho_{(N,ij)}({\cal F}_M(\tau) \Pi_N A\Pi_N)= \frac{1}{2M} \sum_{m= -M}^M
e^{im(\theta_{Ni}-\theta_{Nj} - \tau)} \rho_{(N,ij)}(A)$$
$$= \frac{1}{2M}D_M(\theta_{Ni}-\theta_{Nj} - \tau) \rho_{(N,ij)}(A)$$
where $D_M$ is the Dirichlet kernel $D_M(x) = \frac{sin(M + \frac{1}{2})}{sin(\frac{1}{2})}.$
Hence (4.8) is equivalent to:
$$\lim_{M\rightarrow \infty}\lim_{N \rightarrow \infty}  \frac{1}{d_N}
\sum_{i,j =1}^{d_N} |\frac{1}{2M}D_M(\theta_{Ni}-\theta_{Nj} -
 \tau)|^2 |\rho_{(N,ij)}(A)|^2=0.\leqno(4.9)$$
Given $\epsilon > 0$ we choose $M $ sufficient large so that (4.9) is $\leq \epsilon.$
If we then choose $\delta>0$ so that
$\frac{1}{2M}D_M(x) \geq \frac{1}{2}$ for $x \leq \delta$, the statement of the
theorem follows for $A$ in place of $\sigma \in C^{\infty}({\cal O}).$  

This is actually
the general case: the diagonal part $\oplus_{N=0}^{\infty} \Pi_N A \Pi_N$ of $A$ is its
average relative to $W_t$ and hence its symbol is $S^1$-invariant and may be identified
with a function $\sigma$ on ${\cal O}$.  Since the lower order terms in the symbol make
no contribution in the limit $n \rightarrow \infty$, the statement is only non-trivial for
the Toeplitz multiplier $\Pi \sigma \Pi$. 
 \qed
\medskip

\subsection*{Corollary}{\it The Toeplitz system is quantum weak mixing in the sense that
$$\lim_{N \rightarrow \infty} \omega_N (\hat{A}^*(\chi)\hat{A}(\chi)) = 0$$
for $\chi \not= 1.$}

\subsection*{Proof}: The Szego limit theorem cited above shows that
$$\frac{1}{d_N}||\hat{A}_M^*(\chi)\hat{A}_M(\chi)||_{HS}^2 
=\frac{1}{d_N}\omega_N (\hat{A}^*_M(\chi)\hat{A}_M(\chi)) + o(1) \leqno(4.10).$$
Hence 
$$\frac{1}{d_N}\omega_N (\hat{A}^*_M(\chi)\hat{A}_M(\chi)) \rightarrow 0$$
for $\tau \in \R - 0$ and the Corollary follows from the fact that
$$\omega_N (\hat{A}^*_M(\chi)\hat{A}_M(\chi))\geq 
\omega_N (\hat{A}^*(\chi)\hat{A}(\chi) \leqno(4.11).$$
(For the proof of (4.11) see [Z2,Proposition (1.3iv)].)\qed

\subsection*{Remark}  Although we will not prove it here, the quantum mixing property
${\cal M!}$ is actually equivalent to the weak mixing of $\chi$ on $({\cal O}, \mu).$
The proof is essentially as in Theorem 1 of [Z2], given the modifications above to the
`if' half of that Theorem.  We also refer to [Z2]
for other variants of the weak mixing conditions.  All of these conditions generalize
to the Toeplitz setting and even to the case of essentially general quantized abelian systems.
For the sake of brevity we have only stated the condition which is most concrete in terms
of the eigenfunctions of the system.

\section{Quantized symplectic torus automorphisms: Proof of Theorem D}

In this section, we illustrate the general theory in \S2-4 with the special case of
  quantized symplectic torus automorphisms $g\in Sp(2n, \Z)$.   As will be seen,
if $g$ lies in the theta-subgroup $Sp_{\theta}(2n, \Z)$,  then 
it lifts to a contact transformation $\chi_g$
of the circle bundle  $N_{\Z}/N_{\R}\sim\Hb^{\red}_n/ \bar \Gamma$ over $\R^{2n}/\Z^{2n}$
with respect to the natural contact structure $\alpha$.  Here,
 $N_{\Z}/N_{\R}\sim\Hb^{\red}_n/ \bar \Gamma$ is the quotient
of the Heisenberg group $N_{\R}$ (or reduced Heisenberg group $\Hb^{\red}_n$)
 by its integer lattice $N_{\Z}$ (or reduced lattice $\bar \Gamma$).  
The quantization will then be a
unitary Toeplitz operator of the form $\Pi \chi_g \Pi$, operating on the Hardy space
$H^2_{\Sigma}(N_{\Z}/N_{\R})$ of CR functions  on the quotient.

As mentioned in the introduction, the action of the Toepltitz-quantized torus automorphisms 
on these CR functions will be identified 
with the classical action of the theta group $Sp_{\theta}(2n, \Z)$ 
on the space of theta functions (of variable degree).   The statements in
Theorem D  will follow directly from this link.  To establish it, we will need to draw
on the harmonic analysis of theta functions from [A][A.T], the transformation theory of
theta functions from [K.P],  and the analysis of CR functions on $N_{\Z}/N_{\R}$ from
[F.S][S].  The notational differences between these references explain, and we hope
justify, the notational redundancies in this section.
 
\subsection*{(5.1) Symplectic torus automorphisms}

The starting point is  the affine symplectic manifold
$(T^*\R^n,\sigma )$, where $\sigma  = \Sigma^n_{j=1}dx_j\Lambda
d\xi_j$, and with a co-compact lattice $\Gamma\subset T^*\R^n$ which
we will take to be $\Z^{2n}$.  The quotient $(T^*\R^n/
\Gamma, \sigma )$ is then a symplectic torus.  If $g \in Sp
(T^*\R^n,\sigma )=Sp(2n,\R)$ is a linear symplectic map satisfying
$g(\Z^{2n}) = \Z^{2n}$, then $g$  descends to symplectic
automorphism of the torus (still denoted $g$).

It is convenient to express $g$ in block form
$$g = \begin{pmatrix}A&B\\C&D\end{pmatrix}:\R^n_x\oplus \R^n_\xi
\rightarrow \R^n_x \oplus \R^n_\xi\leqno(5.1.1)$$
relative the the splitting $T^*\R ^n=\R^n_x \oplus \R^n_\xi \simeq 
\R^{2n}$.  Then $g\in Sp(2n,\R)$, i.e.\ $g$ is a symplectic linear map
of $\R^{2n}$, if and only if

\subsection*{(5.1.2)}  (i)  $g^*\in Sp(n,\R)$

(ii)  $A^*C = C^*A, B^*D = D^* B, A^*D - C^* B = I$

(iii)  $AB^* = BA^*, CD^* = DC^*, AD^* - BC^* = I$.

Also $g(\Z^{2n}) = \Z^{2n}$ is equivalent to $a\in
Sp(2n,\Z)$.  ([F, Chapter 4])

\subsection*{(5.2) Kahler and Toeplitz quantization of complex torii}

The quantization of $g$ should be a unitary operator $U_g$ on a
Hilbert space $\Hh$ which quantizes $(T^*\R^n /\Gamma, \sigma
)$.  The method of geometric (Kahler) quantization constructs $\Hh$ as the
space of holomorphic sections of a holomorphic line bundle
$L\rightarrow T^*\R^n/\Gamma$, with respect to a complex
structure $Z$ on $T^*\R^n/\Gamma$.  We will temporarily assume  $Z$ to be the
affine complex structure $J$ coming from the identification
$\R^n_x\oplus \R^n_\xi \rightarrow \C^n ((x,\xi)\mapsto x + i\xi)$.  Later
we will consider more general $Z$.

The line bundle $L$, and its powers $L^{\otimes N}$, are associated
to the so-called prequantum circle bundle $p: X\rightarrow
T^*\R^n/\Gamma$ by the characters $\chi_N$ of $S^1$.  The
definition of prequantum circle bundle also includes a connection
$\alpha $.  As is well-known, in this example $X$ is the compact
nilmanifold $\Hb^{\red}_n/ \bar \Gamma$, where $\Hb^{\red}_n$
is the reduced Heisenberg group $\R^{2n} \times S^1$ and where $\bar \Gamma$ is
a maximal isotropic lattice.
  We pause to
recall the precise definitions, since there are many (equivalent) definitions
of these groups and  lattices.

We will take the group law
of $\Hb^{\red}_n$ in the form
$$(x, \xi, e^{\itt})\cdot (x', \xi', e^{\itt'} ) =
(x + x' ,\xi+\xi', e^{i(t+t'+\half \sigma((x,\xi), (x',\xi')))})\;\leqno(5.2.1)$$
with $ \sigma((x,\xi), (x',\xi')))= \langle \xi, x'\rangle - \langle \xi', x\rangle.$  The
center $Z$ of $\Hb^{\red}_n$ is the circle factor $S^1$.  Evidently, $Z$
acts by left translations on $X$ and its orbits are the fibers of $p$.  The
connection one form is given by
$$\alpha =dt+\half \Sigma^n_{j=1} (x_j d\xi_j - \xi_j dx_j)\;.\leqno(5.2.2)$$

With the group law in the form (5.2.1), the integer lattice $\bar
\Gamma $ is not $\Z ^{2n} \times \{1\}$ (which is not a subgroup)
 but is rather its image under the
splitting homomorphism
$$s : \Z^{2n} \rightarrow \Hb^{\red}_n\;\;\;\;\;\;\; s(m,n):=(m,n, e^{i\pi \langle m, n \rangle}).
\leqno(5.2.3)$$
See subsection (5.8) for the terminology and further discussion.   

Under the action of $Z$, $L^2 (X)$ has the isotypic decomposition
$$L^2 (X) = \bigoplus^\infty_{N=-\infty}H_N$$
where $H_N$ is the set of vectors satisfying $W_t f = e^{2\pi i N t} f$;
here $W_t$ is the unitary representation of $z$ by translations on
$L^2$.  In the standard way, we identify $H_N$ with the sections of
$L^{\otimes N}$.  Thus, $\bigoplus^\infty_{N=0}H_N$ incorporates the
sections of all the bundles $L^{\otimes N}$ at once.

The holomorphic sections of $L^{\otimes N}$ then correspond to the subspace
 $H^2_{\Sigma}(N)$ of
$CR$ functions in $H_N$.  Let us recall the definition  [F.S]: 
First, one defines the left invariant vector fields
\subsection*{(5.2.4a)} \begin{alignat*}{2}
X_j &= \frac{\partial }{\partial x_j} + \xi _j\frac{\partial
}{\partial t}&\qquad (j=1,\ldots, n)&\\
\Xi_j &=  \frac{\partial }{\partial \xi_j}- x _j\frac{\partial
}{\partial t}&\qquad (j=1,\ldots, n)&\\
T&=  \frac{\partial }{\partial t}&\qquad&\end{alignat*}
on $\Hb^{\red}_n$.  They satisfy the commutation relations
$[\Xi_j, X_k] = 2\delta_{jk} T$, all other brackets zero.  Then set
\subsection*{(5.2.4b)} \begin{alignat*}{2}
Z_j&=\frac{\partial }{\partial z_j} + i\bar z_j\frac{\partial
}{\partial t} & = X_j - i \Xi_j&\qquad(j=1,\ldots, n)\\
\bar Z_j&=\frac{\bar\partial }{\partial z_j} - i
z_j\frac{\partial }{\partial t} & = X_j + i \Xi_j&\qquad(j=1,\ldots,
n)\end{alignat*}
(with $Z_j = X_j + i\xi_j)$.  The commutation relations are $[Z_j,
\bar Z_k] = -2i \delta_{jk} T$, all other brackets zero.  One notes
that $\alpha (Z_j) = \alpha (\bar Z_j) = 0\quad (\forall j)$, so the
sub-bundle $T_{1,0}$ of $T(\Hb^{\red}_n)\otimes \C$ defines
a $CR$ structure on $\Hb^{\red}_n$.  The Levi form is given by
$\langle Z_j, Z_k\rangle_L = \half i \langle \alpha , [Z_j,
\bar Z_k]\rangle = \delta_{jk}$, so $\Hb ^{\red}_n$ is
strongly pseudo convex.  All of these structures descend to the
quotient by $\bar \Gamma$ and define a $CR$ structure on $X$. 
The $CR$ functions are the solutions of the Cauchy--Riemann
equations
$$\bar Z_j f = 0\quad\quad (j = 1,\ldots ,n)\;.\leqno(5.2.5)$$

We will denote by $H^2_{\Sigma}(X)$ the $CR$ functions which lie in
$L^2(X)$.
 Under the action of $Z$ we have the isotypic
decomposition
$$H^2_{\Sigma}(X) = \bigoplus^\infty_{N=0} H^2_{\Sigma}(N)\leqno(5.2.6)$$
where  $H^2_{\Sigma}(N):= H^2_{\Sigma} \cap H_N$ is the space of CR vectors transforming by
the $N$-th character $\chi_N$.  Under the identification of
sections of $L^{\otimes N}$ with equivariant functions on $X$ in
$H_N$, the holomorphic sections correspond to $H^2_{\Sigma}(N)([A.T][A][M])$.  As is well-known,
and will be reviewed below,
 the holomorphic sections $\Gamma _{\hol}
(L^{\otimes N})$ are the theta functions of degree $N$.

\subsection*{(5.3) Toeplitz quantization of symplectic torus automorphisms}

Thus far, we have followed the procedure of geometric
quantization theory and have quantized $(T^* \R^n/\Gamma, \sigma
)$ as the sequence of Hilbert spaces $H^2_{\Sigma}(N) \simeq \Gamma_{\hol}
(L^{\otimes N})$.  The next step is to quantize the symplectic map
$g$.  For this, geometric quantization offers no well-defined
procedure in general, and indeed it is  not possible to
quantize general symplectic maps (even very simple ones) in a
systematic way.  In the case of certain $g\in Sp(2n,\Z)$ we can use the Toeplitz method.
These are the elements in the theta-subgroup $Sp_{\theta}(2n,\Z):=\{ g\in Sp(2n,\Z):
AC \equiv 0 (mod 2), BD \equiv 0 (mod 2)\}.$

\subsection*{(5.3.1) Proposition}  {\it Let $g\in Sp_{\theta}(2n, \Z)$,
 and let
$\chi_g: \N_{\R}\rightarrow N_{\R}$ be defined by
$$\chi_g (x, \xi, t) = (g(x,\xi), t)\;.$$
Then $\chi_g$ descends to a contact diffeomorphism of $(X, \alpha )$.}

\begin{pf} First, $\chi_g$ is well-defined on the quotient 
$\Hb^{\red}_n/ \bar \Gamma$ of the Heisenberg group since the elements
of  $Sp_{\theta}(2n,\Z)$ are the automorphisms of $\Hb^{\red}_n$  preserving $\bar \Gamma$.
The last statement follows from the fact that $F(g(m,n))\equiv F(m,n) $ (mod 2)
if $g \in Sp_{\theta}(2n,
\Z)$ and if $F(m,n):= \langle m,n\rangle.$  

 It remains to show that $\chi^*_g\alpha  = \alpha $.  Let us write
$\alpha  = dt +\half (\langle x, d\xi\rangle-\langle \xi,
dx\rangle)$ where $x = (x,\ldots, x_n)$, $\xi = (\xi_1, \cdots,
\xi_n)$ and $\langle a,b\rangle=\Sigma a_i b_i$.  Then $\chi^*_g
\alpha  = dt +\half (\langle x^1, d\xi ^1\rangle-\langle \xi^1,
dx^1\rangle)$ where $x^1 = Ax + B\xi$, $\xi ^1 = Cx + d\xi$ (in the
notation of 3.1-2).  We note that
\begin{align*}
\langle x^1, d\xi^1\rangle - \langle \xi^1, dx^1\rangle&=\langle
(A^* C - C^* A) x, dx\rangle\\
&\quad \quad+\langle (D^* B - B^* D)\xi , d\xi\rangle\\
&\quad \quad+ \langle (D^* A - B^* C)x , d\xi\rangle\tag{5.3.2}\\
&\quad \quad+ \langle (C^* B - A^* D)\xi , dx\rangle\\
&= \langle x, d\xi\rangle - \langle \xi, dx\rangle\end{align*}
by the identities in (5.1.2).  Hence $\chi^*_g \alpha  = \alpha $.
\end{pf}

\subsection*{Remark} Unfortunately, translations $T_{(x_o, \xi_o)}$ on $\R^{2n} / \Z^{2n}$ do
{\it not} lift to contact transformations of this contact structure. They do of
course lift to translations of $X$ by the elements $(x_o, \xi_o,1) \in \Hb^{\red}_n$,
but these do not preserve $\alpha$.  Indeed, $\alpha$ is right-invariant but not bi-invariant
under $\Hb^{\red}_n$, and the invariance was used up in going to the quotient by $\bar \Gamma.$
The only elements of $\Hb^{\red}_n$ which lift to contact transformations are those
which normalize $\bar \Gamma$, namely $N_{\Z}$ itself.
\medskip

As above,  we let $\Sigma = \{ (x, r\alpha_x): x\in X, r>0\}$ denote the
symplectic cone through $(X,\alpha)$ in $T^* X\backslash 0$.  We also let
$\Pi: L^2(X) \rightarrow H^2 (X)$ denote the orthogonal projection (i.e.\
 the Szeg\"o projector) onto the space of $L^2$ CR functions.
  From the analysis of $\Pi$ due to Boutet
de Monvel and Sj\"ostrand [B.S], one knows that $\Pi$ is a
Toeplitz structure on $\Sigma$.  It is obvious that the contact
manifold $(X,\alpha)$ has periodic characteristic  flow (generated by $\Xi =
T$), and that both $\Pi$ and $\chi_g$ commute with $T$.  Hence, by the Unitarization
Lemma,   we can quantize $\chi_g$ as a unitary operator on $H^2 (X)$
of the form
$$U_g = \Pi T_{\chi_{g}} A \Pi \leqno(5..3.3)$$ for some pseudodifferential operator
over $X$ commuting with $T$.  More precisely, it will be unitary if the {\it index} of
$\chi_a$ vanishes, a condition that we will discuss further below.
  Since $U_g$ commutes with $T$, it is diagonal with respect to the
decomposition (5.4) and hence is equivalent to 
sequence of finite rank unitary operators
$$U_{N,g}: H^2_{\Sigma}(N) \rightarrow H^2_{\Sigma}(N)\;,\leqno(5.3.3N)$$
 the finite dimensional quantizations of $g$.

Since the Unitarization Lemma constructs $U_g$ in a canonical fashion from the
contact transformation $\chi_g$, we should be able to determine it completely in
a concrete example.  The first step is to determine the principal symbol, or more
precisely the function given in (3.10).  

To calculate it, we introduce the coordinates $(x, \xi, t, p_x, p_{\xi}, p_t)$ on
$T^*(X)$ with $(x,\xi,t)$ the base coordinates used above and
with   $(p_x, p_{\xi}, p_t)$ the sympletically dual fiber coordinates.  Thus, the
symplectic structure on  $T^*(X)$ is given by 
$$\Omega:= \sum dx_j \wedge dp_{x_j} + d\xi_j \wedge dp_{\xi_j} + dt \wedge dp_t.$$
The cone $\Sigma$ is then parametrized by $i: \R^+ \times X\rightarrow T^*X$, $(r,x,\xi,t) \rightarrow
(x,\xi,t,2r\alpha_x)$ and since this is a diffeomorphism we can use the parameters
as coordinates on $\Sigma.$
The equation of $\Sigma$  is then given by
$$p_x =  r \xi\;\;\;\;\;\;\;\;\;\; p_{\xi} = -  x\;\;\;\;\;\; p_t = -r.$$
Hence,
$$i^*(\Omega) = \sum dx_j\wedge d\xi_j + \alpha \wedge dr.$$

\medskip

We recall that $\Sigma$ is the characteristic variety of the involutive system (5.2.5)
and that the symbol $\sigma_{\Pi}$ of the Szego projector involves the positive
Lagrangean sub-bundle (3.5) of $T\Sigma^{\bot}\otimes \C$.  We now describe these
objects concretely:

\subsection*{(5.3.4) Proposition}{\it At a point $p= i(x_o, \xi_o, t_o, r_o) \in \Sigma,$
we have:

(a)
$$T_p\Sigma^{\bot}   = sp \{ X_j + r_o \frac{\partial}{\partial p_{x_j}}, \Xi_j - r_o
\frac{\partial}{\partial p_{\xi_j}}\}$$

(b)
$$\Lambda_p = sp_{\C}\{\overline{Z}_j + r_o (\frac{\partial}{\partial p_{x_j}} + i
\frac{\partial}{\partial p_{\xi_j}}) \}$$}

\subsection*{Proof} 

(a) Using the above parametrization, we find that
$$\begin{array}{l} i_{*} \frac{\partial}{\partial x_j} = \frac{\partial}{\partial x_j} -
r_o \frac{\partial}{\partial p_{\xi_j}}\\
i_{*}\frac{\partial}{\partial \xi_j}=\frac{\partial}{\partial \xi_j} + r_o
\frac{\partial}{\partial p_{x_j}}\\
i_{*} \frac{\partial}{\partial t} = \frac{\partial}{\partial t}\\
i_{*} \frac{\partial}{\partial r}=  \xi_{oj} \frac{\partial}{\partial p_{x_j}}
-  x_{oj} \frac{\partial}{\partial p_{\xi_j}} -\frac{\partial}{\partial p_t} \end{array}$$
from which it is simple to determine the vectors $X$ such that $\Omega(X, T_p\Sigma)
=0.$ 

(b) The operators $D_j$ of \S3 are the operators $\overline{Z}_j$
of (5.4b) whose symbols are given by
$$\sigma_{D_j}= i p_{x_j} - p_{\xi_j} +(x_j + i\xi_j)p_t.$$
Their Hamilton vector fields
$$H_{\sigma_j} = \frac{1}{i}(X_j + i \Xi_j + ir_o ( \frac{\partial}{\partial p_{x_j}} + i
\frac{\partial}{\partial p_{\xi_j}})) $$
 are easily seen to agree (up to complex scalars) with
the vector fields asserted to span $\Lambda_p.$ \qed

We now wish to determine the vacuum states corresponding to $\Lambda$ and
$\chi(\Lambda)$.  Recall that,
given a symplectic frame of $T_p\Sigma^{\bot}$, we get a representation $d\rho_p$ of the
Heisenberg algebra on the space ${\cal S}((\Sigma^{\bot})_p$ (see \S3) and that the vacuum
state $e_{\Lambda_p}$ is the unique state annihilated by the elements of $\Lambda_p$.
To determine it, we choose the symplectic frame 
$${\cal B}_p:= \{\frac{1}{\sqrt{r_o}}Re H_{\sigma_j}, 
\frac{1}{\sqrt{r_o}}Im  H_{\sigma_j}, j= 1,\dots,n\}$$
and write a vector $V \in (T_p\Sigma)^{\bot}$ as $V=\sum \alpha_j 
\frac{1}{\sqrt{r_o}}Re H_{\sigma_j} + \beta_j\frac{1}{\sqrt{r_o}}Im H_{\sigma_j} .$
We observe that $\{\frac{1}{\sqrt{r_o}}Re H_{\sigma_j},
\frac{1}{\sqrt{r_o}}Im  H_{\sigma_j},T\}$ form a Heisenberg algebra and that under 
the Schrodinger representation $d\rho_p$ they
go over to $\{\frac{\partial}{\partial \alpha_j}, \alpha_j, 1\}$.  

\subsection*{(5.3.5) Proposition}{\it With the above notation: The vacuum state
$e_{\Lambda_p}$ equals the Gaussian $e^{-\half |\alpha|^2}$.}

\subsection*{Proof} The annihilation operators in the representation $d\rho_p$ are
given by the usual expressions $\frac{\partial}{\partial \alpha_j} + \alpha_j$ and
hence the vacuum state is the usual one in the Schrodinger representation.\qed

Now consider the image of $\Lambda $ under the contact transformation
$\chi_g$, or more precisely its lift as the symplectic transformation 
$$\tilde{\chi}_g(x,\xi,t,p_x,p_{\xi},p_t)= (A x + B\xi, Cx + D\xi, t,  D p_x
 -C p_{\xi}, -B p_x + A p_{\xi}, p_t) \leqno(5.3.6)$$
of $T^*X$.
Of course, it is linear in the given coordinates.  We would like to compare $d\tilde{\chi}_{g,p}
(\Lambda_p)$ and $\Lambda_{\tilde{\chi}_g(p)}.$

\subsection*{(5.3.7) Proposition}{\it Under $d\tilde{\chi}_{g,p}$ we have, in an obvious
matrix notation:

(a)$$ \begin{array}{l}X \rightarrow A X + C \Xi \\ \Xi_j \rightarrow B X + D\Xi\end{array}$$

(b)$$ \begin{array}{l}\frac{\partial}{\partial {p_x}}\rightarrow B\frac{\partial}{\partial p_{x}} +
D\frac{\partial}{\partial p_{\xi}}\\
\frac{\partial}{\partial p_{\xi }}\rightarrow A\frac{\partial}{\partial p_{x}} +
C\frac{\partial}{\partial p_{\xi}}\end{array}$$

(c) $$ \begin{array}{l} Re i H_{\sigma} \rightarrow A (Re i H_{\sigma}) + C( Im i H_{\sigma})\\
 Im i H_{\sigma} \rightarrow  B (Re i H_{\sigma}) + D( Im i H_{\sigma})\end{array}$$

(d) $d\tilde{\chi}_{g,p}{\cal B}_{p} = g^{*} {\cal B}_{\chi(p)}.$

(e) $e_{\Lambda_{\chi_g}} = \mu(g^{*}) e_{\Lambda}$ where $\mu$ is the metaplectic
representation. }

\subsection*{Proof}  The formulae in (a)-(b) are  easy calculations left to the reader.
The ones in (c)-(d) are  immediate consequences.  The statement in (e) follows from
the change in the Schrodinger representation under a change of metaplectic basis [B.G].
\qed

The desired principal symbol is determined by the following proposition. 

\subsection*{(5.3.8) Proposition}{\it  Let $g= \left(\begin{array}{ll} A & B\\C & D \end{array}
\right)$.  Then
 the inner product $\langle e_{\Lambda_{\chi_g}},
e_{\Lambda}\rangle$  in the Schrodinger representation equals: 
$$\langle e_{\Lambda_{\chi_g}},e_{\Lambda}\rangle=
2^{\frac{n}{2}} (det(A + D + iB -iC))^{-\half}.$$}

\subsection*{Proof} Let $Z = X + i Y$ be a complex symmetric matrix with $Y>>0$,
and let $\gamma_{Z}(x):= e^{\frac{i}{2}  <Zx,x>}$ be the associated Gaussian.  The action of
an element $g \in Mp(2n, \R)$ is the given by:
$$\mu(g^{ * -1})\gamma_{Z} = m(g,Z) \gamma_{\alpha(g)Z}$$
where
$$m(g,Z) = det^{-\half}(CZ +D),\;\;\;\;\;\;\alpha(g)Z = (AZ+B)(CZ+D)^{-1}$$
(see [F, Ch.4.5]).  We may assume $e_{\Lambda} = \frac{\gamma_i}{||\gamma_i||}$ and
since
$$\mu(g^*)\gamma_i = m(g^{-1}, i) \gamma_{g^{-1} i}$$ 
we have
$$\langle e_{\Lambda_{\chi_g}},e_{\Lambda}\rangle =
m(g^{-1},i)  \langle \gamma_{g^{-1}i},  \gamma_i \rangle.$$ 
The inner product of two  Gaussians is given by
$$\langle \gamma_{\tau}, \gamma_{\tau'}\rangle =
\int_{\R^n} e^{\frac{i}{2} \langle (\tau - \overline{\tau'}) \xi, \xi\rangle} d\xi=
\frac{1}{\sqrt{det [-i (\tau - \overline{\tau'})]}} \leqno(5.3.9)$$
with the usual analytic continuation of the square root [F].  Putting $\tau = g^{-1}iI$ and
$\tau' =  iI$ and simplifying we get the stated formula.\qed
\medskip

  For future reference we will rephrase the
previous proposition in the following form:

\subsection*{(5.3.10) Corollary}{\it The Toeplitz operator
$$U_{g}:= m(g) \Pi \chi_g \Pi\;\;\;\;\;\;\;\;\;\;\;\; 
m(g)=2^{\frac{-n}{2}} (det(A + D + iB -iC))^{\half}$$
is unitary modulo compact operators.}

We will now see that $U_{g}$ is actually unitary if $g \in Sp_{\theta}(2,\Z)$ or
if $g$ lies in the image of the natural embedding of $Sp_{\theta}(2,\Z)$ in
 $Sp_{\theta}(2n,\Z)$. The same
statements are  true for the other elements $Sp_{\theta}(2n,\Z)$, but we will restrict
to these elements so that we can easily quote from [K.P].
 
 \subsection*{(5.4) Theta functions }
\medskip

 We begin with a rapid review of  the transformation theory of theta functions
under elements $g \in Sp_{\theta}(2,\Z)$.
As above, in  dimensions larger than two,
 $Sp_{\theta}(2,\Z)$ is understood to be embedded in $Sp_{\theta}(2n,\Z)$ as 
the block matrices
$(\left( \begin{array}{ll} a I_n & b I_n \\ c I_n & d I_n \end{array}\right)$ with
$I_n$ the $n\times n$ identity matrix.   For this case, we 
 closely follow the exposition of Kac-Peterson [KP].
 For more classical treatments of
transformation laws, and in
 more general cases, see [Bai][Kloo].  

Notation: ${\cal H}_+:= \{ \tau = x + iy | x,y \in \R, y>0\}$ will denote the Poincare
upper half-plane and the standard action of $SL(2,\R)$ on ${\cal H}_+$ will be written
$$\left( \begin{array}{ll} a & b \\ c & d \end{array}\right) \tau = \frac{a \tau + b}
{c \tau + d}.$$
 $U_{\R}\equiv \R^n$ will denote a real vector space of dimension $n$, equipped
with a positive definite symmetric bilinear form $<,>$, and  $U=U_{\R}\otimes \C.$
The Heisenberg group will be taken in the unreduced form $N_{\R} =
U_{\R}\times U_{\R} \times \R$
with multiplication $(x,\xi, t)(x', \xi', t')= (x+x', \xi + \xi', t + t' + \frac{1}{2}
[\langle x',\xi\rangle  -\langle x, \xi'\rangle]).$

To quantize  $SL(2, \Z)$ as a group action, one has to lift to the metaplectic group
$$Mp(2, \R) := \{(\left( \begin{array}{ll} a & b \\ c & d \end{array}\right), j):
j(\tau)^2 = c\tau + d, j: {\cal H}_+ \rightarrow \C \;\;\;\;\;\;\;\mbox{holomorphic}\}.$$
Set
$$Y:= {\cal H}_+ \times U \times \C$$
and let $Mp(2, \R)$ act on $Y$ by
$$(\left( \begin{array}{ll} a & b \\ c & d \end{array}\right), j) (\tau, z, t): =
( \frac{a \tau + b}{c \tau + d}, \frac{z}{c \tau + d}, t + \frac{c}{2}  \frac{<z,z>}
{c \tau + d}).\leqno(5.4.1)$$
Also let the Heisenberg group act by
$$(x,\xi, t_o) \cdot (\tau, z, t):= (\tau, z - x + \tau \xi, t - <\xi, z> - \frac{1}{2}\tau <\xi,\xi>
+ \frac{1}{2} <x,\xi> + t_o).\leqno(5.4.2)$$
Let $G_{\R}$ be the semi-direct product of $Mp(2,\R)$ with  $N_{\R}$, with
$gng^{-1}= g\cdot n$.  It  acts on functions on $Y$ by
$$f|_{(\left( \begin{array}{ll} a & b \\ c & d \end{array}\right), j)}(\tau,z,t)=
j(\tau)^{-n} f(\left( \begin{array}{ll} a & b \\ c & d \end{array}\right)(\tau,z,t))\leqno(5.4.3)$$
$$f|_n(\tau,z,t)= f(n(\tau,z,t)).$$
Now let $L$ denote a lattice of full rank in $U_{\R}$ such that
 $\langle \gamma, \gamma'\rangle
\in \Z$ for all $\gamma,\gamma' \in L$, let $L^*$ be the dual lattice $\{\gamma: 
\langle \alpha, \gamma\rangle
 \in \Z (\forall \alpha \in L)\}$.  For the sake of simplicity we will
assume $L=L^*$ and in fact that $L= \Z^{n}.$  Then define the integral subgroup
 $$N_{\Z} = \{(x,\xi,t) \in N_{\R}: x,\xi \in L, t + \half \langle x,\xi\rangle
 \in \Z\}.\leqno(5.4.4)$$
The normalizer $G_{\Z}$ of $N_{\Z}$ in $N_{\R}$ is given by
$$G_{\Z} = \begin{array}{l}\left\{ \left(\left(\begin{array}{ll}
 a & b \\ c & d \end{array}\right), j\right) (\alpha,\beta, t) \in G_{\R} : 
\left( \begin{array}{ll} a & b \\ c & d\end{array} \right) \in SL(2,\Z);\right.\\
\left. bd\langle \gamma, \gamma\rangle  \equiv 2 \langle \alpha, \gamma \rangle mod 2\Z,
ac\langle \gamma, \gamma\rangle \equiv 2 \langle \beta, \gamma \rangle mod 2 \Z,
\forall \gamma \in \Z^n \right\}\end{array} .\leqno (5.4.5)$$
In particular, $Sp_{\theta}(2, \Z) \subset G_{\Z}.$

\subsection*{(5.4.4) Definition}{\it  The space of theta functions of degree N 
is the space $\tilde{Th}_N$ of holomorphic functions $f$ on $Y$ satisfying:
$$f|_n = f\;\;\;\;(\forall n \in N_{\Z}),  \;\;\;\;\;\;\;\;\;\;f|_{(o,o,t)} = e^{-2\pi i N t} f.$$
The entire ring of theta functions is the space
$$\tilde{Th} := \bigoplus_{N \in {\bf N}} \tilde{Th}_N.$$}
\medskip

We observe that $\tilde{Th}_1$ puts together the  holomorphic (pre-quantum) line
bundles $${\cal L}_{\tau} \rightarrow U/(L + \tau L) \leqno(5.4.5)$$
 over $\R^{2n}/\Z^{2n}$ -torus as the
one-parameter family of complex structures parametrized by $\tau$ varies.  Indeed,
following [KP, p.181] we observe that the natural projection
$$\pi : Y={\cal H}_+ \times U \times \C \rightarrow {\cal H}_+ \times U$$
defines a holomorphic line bundle.  The group $\overline{N}_{\Z}:=N_{\Z}/ \Z$ acts freely by
bundle maps, so the  quotient line bundle
$$\overline{\pi}: Y / \overline{N}_{\Z} \rightarrow ({\cal H}_+ 
\times U)/\overline{N}_{\Z}\leqno(5.4.6)$$
defines a holomorphic line bundle which for each fixed $\tau$ restricts to $(5.4.6 \tau)$.
Similarly for the powers ${\cal L}^{\otimes N}$.   Hence $\tilde{Th}$ simeltaneously
puts together theta-functions of all degrees and complex structures in the one-parameter
family above.  If we fix $\tau$ we get the space $Th_N^{\tau}$ of holomorphic sections
of  ${\cal L}_{\tau}^{\otimes N}$, that is, the space 
 of holomorphic theta functions of degree N relative to $\tau.$
\medskip

\subsection*{(5.5) Classical theta functions of degree N and characteristic $\mu$ a la [K.P]}
 \medskip

We now introduce a specific basis of the theta functions of any degree and with
respect to any complex structure $\tau$.  These are not yet the theta functions which
will play the key role in Theorem D, but are a preliminary version of them.  We follow
the notation and terminology of [K.P] except that we put $L = \Z^{n} = L^*$.

  For $\mu \in \Z^n/ N\Z^n$, define
the {\it classical theta function of degree n} and {\it
characteristic $\mu$} with respect to the complex structure $\tau$ by: 
$$\Theta_{\mu,N}(\tau,z,t):= e^{-2 \pi i N  t}\sum_{\gamma \in \Z^{n} + \frac{\mu}{N}}
e^{ 2\pi i N  \{\half \tau <\gamma,\gamma>  -  <\gamma, z>\}}.\leqno(5.5.1)$$

When the degree N=1 and $\mu=0$ this is the {\it Riemann theta function} for the
lattice $\Z^n$,
$$\Theta(\tau, z, t): = e^{-2 \pi i t}\sum_{\gamma \in \Z^n} e^{  2\pi i \{\half
\tau <\gamma,\gamma>
 -  <\gamma,z>\}}\leqno(5.5.2)$$
while the general theta function of degree 1 and characteristic $\mu \in \R^{n}/\Z^{n}$
 is given by
$$\Theta_{\mu} (\tau, z, t):= \Theta |_{(0,-\mu,0)} =e^{-2 \pi i t}
\sum_{\gamma \in \Z^n + \mu} e^{ 2\pi i \{\half \tau <\gamma,\gamma> -  <\gamma,z>\}}
.\leqno(5.5.2 \mu)$$

  One has the following:
\medskip

\subsection*{(5.5.3) Proposition}(see [K.P., Lemma 3.12]) {\it Fix $\tau$.
  Then:
$$\{ \Theta_{\mu,N}|_{Y_{\tau}} \}_{ \mu \in \Z^n/N \Z^n} \mbox{ is a
$\C$-basis of}\;\;Th_N^{\tau}.$$}

\medskip

\subsection*{(5.6) Transformation laws}
\medskip

The transformation laws for  classical theta functions are given by the following:

\subsection*{(5.6.1) Transformation law ([K.P., Proposition 3.17]}{\it Let 
$g=(\left( \begin{array}{ll} a & b \\ c & d \end{array}\right),j) \in Mp(2,\R)$
be an element satisfying:
$$Nbd \equiv 0 \mbox{mod}\;\;\;2\Z\;\;\;\;\;\;\;\;\;\;\;Nac \equiv 0 \mbox{mod}\;\;2Z.$$
 Then there exists $\nu(n,g) \in \C$ such that, for $\mu \in \Z$,
$$\Theta^L_{\mu,n}|_g =\nu (n,g) \sum_{{\alpha \in \Z^n}\atop{c \alpha\;\; mod\;\;
N \Z^n}} e^{i \pi [N^{-1} cd \alpha^2 + 2 N^{-1}bc \alpha \mu + N^{-1}ab \mu^2]}
\Theta^L_{a \mu + c \alpha, N}.$$
The matrix  of $g$ with respect to the above $\C$-basis is unitary.}
\medskip

The multiplier $\nu(n,g)$ is described in detail in [KP, loc.cit] and involves the
Jacobi symbol.

For the generators 
$$ S= \left( \begin{array}{ll} 0 & -1 \\ 1 & 0 \end{array}\right),\;\;\;\;\;
T= \left( \begin{array}{ll} 1 & 1 \\ 0 & 1 \end{array}\right)$$
of  $SL(2, \Z)$, and for  n=1,
 the transformation law reads:

$$\Theta_{\mu,N} (-\frac{1}{\tau}, \frac{z}{\tau}, t + \frac{z^2}{2\tau}) =
(-i\tau)^{\frac{1}{2}} (N)^{-\frac{1}{2}} \sum_{\alpha \in \Z / N \Z}
e^{-\frac{2 \pi i\mu \alpha}{N}} \Theta_{\alpha, N}(\tau, z, t) \leqno(5.6.2 S)$$

$$\Theta_{\mu,N}(\tau + 1, z, t) = e^{2 \pi i\frac{|\mu|^2}{N}} \Theta_{\mu,N}(\tau,z,t).
\leqno (5.6.2T)$$

We note that (5.6.2S) is the formula for the finite Fourier transform on $\Z/N$ [A.T., p.853].
\medskip

\subsection*{(5.7) The space $\Theta^{\tau}(N)$  of theta functions
 $\vartheta^{\tau}_{\mu,N}$}
\medskip

We now specify the theta functions which will play the key role in 
linking the classical transformation
theory to the action of the quantized contact transformation $U_{\chi_g}$.  They
are essentially the (variable degree) versions of the `most natural and basic' theta
functions of [M] and coincide with the span $\Theta(N)$ of the theta-functions
 denoted $\phi_{Nj}$ in [A.T].  

The reader should note that the expression $\Theta_{\mu,N} |_n (\tau, z, t)$ depends
on many variables.  In different articles, different sets of variables are viewed as
the significant ones.  Here we wish to regard theta-functions as functions on 
$N_{\Z}/ N_{\R}$ so we emphasize the $n \in N_{\R}$ variable.  In other contexts,
$(\tau,z)$ are viewed as the significant variables (cf. [Bai][M][Kloo]).

\noindent{\bf (5.7.1) Definition} {\it  For $\mu \in \Z^n/N \Z^n$,  put:
$$\vartheta^{\tau}_{\mu,N} (x, \xi, t): = e^{-2 \pi i N t}
\Theta_{\mu,N} |_{(x, \xi,0)}(\tau,0,0)$$
$$ =e^{- 2\pi i Nt}\sum_{\gamma \in \Z^n} 
e^{ 2 \pi i N [  \frac{\tau}{2} 
\langle \xi + \frac{\mu}{N}+ \gamma, \xi +\frac{\mu}{N} + \gamma \rangle + \langle 
\frac{\mu}{N} \xi +\gamma, x\rangle]}. $$}

The significance of these theta-functions is due to the following:

\subsection*{(5.7.2) Proposition}{\it The theta functions $\vartheta^{\tau}_{\mu,N}$ satisfy:

\noindent(i) $\vartheta^{\tau}_{\mu,N}\in H_N(N_{\Z}\backslash N_{\R})$;
 
\noindent(ii)  As $\mu$ runs over $\Z^n / N \Z^n$, $\vartheta^{\tau}_{\mu,N} (x, \xi, t)$ forms
  a basis of the  CR functions of degree (= weight) N on $N_{\Z}\backslash N_{\R}$.}

\subsection*{Proof} 

 First, for $\mu \in \Z^n/ N\Z^n, \Theta_{\mu,N}$ is $N_{\Z}$-invariant as a function on
$Y$, that is, $\Theta_{\mu,N} |_{n} = \Theta_{\mu,N}$ for $n \in N_{\Z}$ ([K.P., 3.2]). 
It follows that
$$\vartheta_{\mu,N}^{\tau} (n (x,\xi,0))  = \Theta_{\mu,N} |_{n (x,\xi,0)}(\tau,0,0)
= \Theta_{\mu,N} |_{(x, \xi,0)}(\tau,0,0)$$
since $|_n$ is a right action.

The CR property is a direct consequence of the fact that the theta functions are
holomorphic on $Y$.  To give a compete proof of this, one would have to introduce
the  CR structure $\overline{Z}_j$  on $N_{\Z}\backslash N_{\R}$ corresponding 
to a complex structure $Z$ on
the torus $N_{\Z}\backslash N_{\R}/ {\cal Z}$ (with ${\cal Z}$ the center), and verify
that differentiation of $\Theta_{\mu,N} |_{(x, -\xi,0)}(\tau,0,0)$ by $\overline{Z}_j$ in the
$(x, \xi,t)$-variables is equivalent to differentation of $\Theta_{\mu,N}(\tau,z,t)$
in  $\overline{\partial_z}$.  For the details of this calculation we
 refer the reader to [M, p.22] or [A].

Granted the CR property, the statement that the $\vartheta^{\tau}_{\mu,N}$'s form
a basis for the CR funtions of weight N relative to the CR structure $\tau$ follows from
Proposition (5.5.3). \qed
\medskip

The proposition has the following 
representation-theoretic interpretation:  $H_N(N_{\Z}\backslash N_{\R})$ is reducible
as a unitary representation of $N_{\R}$ for $N>1$, and the space $H^2_{\Sigma}(N)$ of
 CR functions in $H_N$
consists of the lowest weight vectors.  For the multiplicity theory, see  [A][A.T].

We now record the modified transformation laws for the theta functions
 $\vartheta^{\tau}_{\mu, N}$ under elements $g \in Sp_{\theta}(2n, \Z)$.  It will be
these transformation laws which will be used to prove Theorem D.
\medskip

\subsection*{(5.7.3) Proposition (Transformation laws for $\vartheta^{\tau}_{\mu,N}$) }
{\it As above, let
$g=(\left( \begin{array}{ll} a & b \\ c & d \end{array}\right),j) \in Mp(2,\R)$
be an element satisfying
$$Nbd \equiv 0 \mbox{mod}\;\;\;2\Z\;\;\;\;\;\;\;\;\;\;\;Nac \equiv 0 \mbox{mod}\;\;2Z.$$
 Then there exists $\nu(N,g) \in \C$ such that,
 for $\mu \in \Z^n / N\Z^n$,
$$\vartheta^{\tau}_{\mu,N}(g\cdot (x,\xi, t)) =\nu (N,g) j(g^{-1}\tau)^n
\sum_{{\alpha \in \Z^n}\atop{c \alpha mod
N\Z^n}} e^{i \pi [N^{-1} (cd) \alpha^2 + 2 N^{-1}bc \alpha \mu + N^{-1}(ab) \mu^2]}
\vartheta^{\tau'}_{a\mu - c\alpha,N}((x,\xi,t)$$
with $\tau' = g^{-1}\tau =\frac{d\tau -b}{-c\tau + a}.$
The matrix  of $g$ with respect to the above $\C$-basis is unitary.}

\subsection*{Proof} We may (and will) set $t=0$ on both sides.  Then,
$$\vartheta^{\tau}_{\mu,N} (ax + b\xi, cx + d\xi,0):=
\Theta_{\mu,N} |_{(ax + b\xi, cx + d\xi,0)} (\tau,0,0)
=\Theta_{\mu,N} |_{g\cdot (x, \xi,0)}(\tau,0,0).$$
  Here,
$g \cdot (x,\xi,0)$ is the action of $(g, j) \in Mp(2,\R)$ on $(x, \xi, 0)$ as
an automorphism of $N_{\R}$.

 Now recall that in  the semi-direct
product $Mp(2,\R)N_{\R}$, we have $(g,j)n(g,j)^{-1}  =( g\cdot n).$
Hence
$$\Theta_{\mu,N} |_{g\cdot (x, \xi,0)}(\tau,0,0)=
\Theta_{\mu,N} |_{(g,j) (x, \xi,0)(g,j){ -1}}\cdot(\tau,0,0))=
j(g^{-1}\tau)^n[\Theta_{\mu,N} |_{(g,j)}] |_{(x, \xi,0)} ( g^{ -1} \tau,0,0).\leqno(5.7.4)$$
Applying the transformation laws (5.6.1), the last expression becomes
$$=\nu (N,g)j(g^{-1}\tau)^n \sum_{{\alpha \in \Z^n}\atop{c \alpha \;mod\;
N\Z^n}} e^{i \pi [N^{-1} (cd) \alpha^2 + 2 N^{-1}bc \alpha \mu + N^{-1}(ab) \mu^2]}
\Theta_{a \mu - c \alpha, N}|_{(x, \xi,0)} (g^{-1}\tau, 0, 0)$$
$$=\nu (N,g) j(g^{-1}\tau)^n\sum_{{\alpha \in \Z^n}\atop{c \alpha \;mod\;
N\Z^n}} e^{i \pi [N^{-1} (cd) \alpha^2 + 2 N^{-1}bc \alpha \mu + N^{-1}(ab) \mu^2]}
\vartheta^{\tau'}_{a \mu - c \alpha, N} (x, \xi,0).$$
The unitarity of the matrix of coefficients follows from the usual transformation rule.
 \qed

\subsection*{(5.8) $\Theta_N^{\tau}$ as a $Heis(\Z^n/N)$-module}

As mentioned above, $\Theta_N^{\tau}$ is an irreducible representation for the
finite Heisenberg group $Heis(\Z^n/N)$.  We pause to define this group and its action
on $\Theta_N^{\tau}$. This will clarify the distinguished role of the classical
theta functions as a basis for $\Theta_N^{\tau}$ and hence will make explicit
the isomorphism to $L^2(\Z/N)$, which is the setting for the quantized cat maps
in [H.B][dE.G.I][Kea].  It will also clarify the relation between the dynamics of cat maps as studied
in the semi-classical literature and those studied in [B][B.N.S].

In the following, $\C_1^*$ denotes the unit circle in $\C$, 
$\C_1^*(N)$ denotes the group of Nth roots of unity, and
$\pm C_1^*(N)$ denotes the group of elements $\pm e^{2\pi i \frac{j}{N}}$.
\medskip

\noindent{\bf (5.8.1) Definition}{\it The finite Heisenberg group $Heis(\Z^n/N)$ is the
subset of elements of
$$\Z^n / N \Z^n \times \Z^n / N \Z^n \times (\pm \C_1^*(N)) $$
generated by $\Z^n/N \times \Z^n/N$ and $\C_1^*(N)$ under
 the group law
$$(m,n, e^{i\phi}) \cdot (m', n', e^{i\phi'}) = (m+m', n+ n', e^{\frac{2\pi i}{N} \sigma((m,n), (m',n'))} 
e^{i (\phi+ \phi')}).$$}

In the terminology of [M], $Heis(\Z^n/N)$ is a generalized Heisenberg group in the
following sense: A general Heisenberg group $G = Heis(K, \psi)$ is
 a central extension by  $\C_1^*$ of a locally
compact abelian group $K$:
$$1 \rightarrow \C_1^* \rightarrow G \rightarrow K \rightarrow 0\leqno(5.8.2)$$
satisfying the following conditions:
\medskip

(i) As a set $G =  K \times \C_1^* $;

(ii) The group law is given by 
$$(x, \lambda) \cdot (\mu, y) = (\lambda \mu \psi(x,y), x+ y)$$
where $\psi: K \times K \rightarrow\C_1^*$ is a 2-cocycle:
$$\psi(x,y)\psi(x+y,z) = \psi(x, y+z)\psi(y,z);$$

(iii) Define a map $e: K \times K \rightarrow \C_1^*$  by
$$e(x,y) = \tilde{x}\tilde{y}\tilde{x}^{-1}\tilde{y}^{-1}$$
where $\tilde{x},\tilde{y}$ are any lifts of $x,y$ to $G$ ($e(x,y)$ is independent of the
choice).  Also define $\phi: K \rightarrow \hat{K}$ by $\phi(x)(y)= e(x,y)$. Then
$\phi$ is an isomorphism.  Here, $\hat{K}$ is the group of characters of $K$.

In the case of $Heis(\Z^n /N)$, $K= \Z^n /N \Z^n \times \Z^n / N\Z^n$ and $\psi$ is
given by $\psi(v,w):= e^{\frac{2\pi i}{N} \sigma(v,w)}$ where $\sigma$ is the restriction of the
symplectic form to $\Z^{2n}$.  Also, we consider the finite subgroup generated by
$K$ and by $\C_1^*(N).$
  
The analogues of Lagrangen subspaces in the case $K=T^*\R^n$ are the maximally
 isotropic subgroups.
Here, a subgroup $H\subset K$ is called {\it isotropic} if $e_{H\times H} \equiv 1$ and
is {\it maximally isotropic} if it is maximal with this property.  
Examples of maximal isotropic subgroups of $Heis(\Z^n/N)$ are given by  $\Z^n / N\Z^n$
 and  by the character group $\widehat{\Z^n / N\Z^n}$.

 Given any isotropic
subgroup, there is a (splitting) homomorphism
$$s : H \rightarrow G\;\;\;\;\;\;\;\;\;\;\;\;\;\;\;\;\;\;\;\;\;\;
s(h) = (h, F(h))\leqno(5.8.3)$$
such that $\pi \cdot s = id_H$.  Here,
$\pi: G \rightarrow K$ is the map in (5.8.2). The map  $F(h)$ satisfies:
$$\frac{F(a + b)}{F(a) \cdot F(b)} = \psi(a,b)\;\;\;\;\;\;\;\;\;\;(a,b \in H).$$

Given a maximal isotropic subgroup $H \subset K$ and a splitting homomorphism
$s: H \rightarrow G$, one defines the Hilbert space:
$${\cal H} = \{ f: K \rightarrow {\bf C}: f \in L^2(K/H), f(x+ h) = F(h)^{-1} \psi(h,x)^{-1}
f(x) \;\;\forall h \in H\} \leqno(5.8.4)$$
and the representation $\rho$ of $G$ on ${\cal H}$ 
$$\rho(k, \lambda) f (x) := \lambda \psi(x,k) f(x + k).$$
Then: $\rho$ is an irreducible representation, and is the unique irreducible with the given
central character.

The choice of $\widehat{\Z^n / N\Z^n}$
  gives a close analogue of the Schrodinger representation
in the real case.  The associated Hilbert space may be identified with $L^2(\Z^n / N\Z^n)$
and the representation is given by
$$U_{(0,0 \lambda)} f(b) = \lambda f(b)\;\;\;\;\;\;\;U_{(a, 0,0)}f(b)= e^{\frac{2\pi i}{N}
\langle a, b \rangle} f(b)
\;\;\;\;\;U_{(0, \chi_a, 0)} = f(b + a). \leqno(5.8.5)$$
\medskip

Now let us return to $\Theta^{\tau}_N$. We first observe that $\Theta_{\mu, N}$
is constructed from $\Theta$ by:
$$\begin{array}{l}\Theta_{\mu, N}(\tau, z, t) = \Theta^{N}_{\frac{\mu}{M}} (\tau, z, Nt)\\
\Theta_{\mu} (\tau, z, t)= \Theta |_{(o, -\mu,o)}(\tau,z,t)\end{array}\leqno(5.8.6)$$
 where
 $\Theta^{N}$ is the same as $\Theta$ except that the complex quadratic form
$<\cdot ,\cdot >$ is replaced by $N <\cdot ,\cdot >.$  

We further observe with [K.P (3.10)] that

$$\begin{array}{l}\Theta_{\mu,N} |_{(\frac{\mu'}{N},0,0)}=
 e^{\frac{2\pi i}{N} <\mu,\mu'>} \Theta_{\mu,N}\\
\Theta_{\mu,N} |_{(0,\frac{\mu'}{N},0)}=  \Theta^L_{\mu-\mu',N}\end{array}\leqno(5.8.7)$$
where $\mu, \mu' \in $ vary over $\Z^n / N\Z^n.$  Since $\Theta_{\mu,N}$ depends only
on $\mu$ mod $N\Z^n$, we see that (5.8.7) defines the same representation of
$Heis(\Z^n /N)$ as in (5.8.5).

The same situation holds  for the theta functions $\vartheta_{\mu, N}^{\tau}$, but we
rephrase things slightly.  First, with [A.T]  let us set
$$\vartheta^{\tau,N} (x,\xi,t):= e^{- 2\pi i t}\sum_{\gamma \in \Z^n} 
e^{ 2 \pi i  [N  \frac{\tau}{2} 
\langle \xi + \gamma, \xi + \gamma \rangle + \langle \gamma, x\rangle]} \leqno(5.8.7).$$
One can verify that $\vartheta^{\tau, 1} = \vartheta^{\tau}_{0, 1}.$  Then define the
Heisenberg  dilations
$$D_m : N_{\R} \rightarrow N_{\R}\;\;\;\;\;\;\;\;\;D_m(x,\xi,t) = (mx, \xi, mt)
\leqno(5.8.8)$$
which are automorphisms of $N_{\R}$.  Associated to them are the dilation
operators
$$D_N: \Theta_m \rightarrow \Theta_{Nm}\;\;\;\;\;\;\;D_N f = f \cdot D_N.$$
Then we have
$$D_N \vartheta^{\tau, N} = \vartheta^{\tau}_{0, N} 
=e^{- 2\pi i N t}\sum_{\gamma \in \Z^n} 
e^{ 2 \pi i N  [  \frac{\tau}{2} 
\langle \xi + \gamma, \xi + \gamma \rangle + \langle \gamma, x\rangle]} \leqno(5.8.9)$$
From (5.8.7) we conclude that 
$$\vartheta^{\tau}_{\mu,N} =\vartheta^{\tau}_{0,N} |_{(0,-\frac{\mu}{N},0)}=
e^{- 2\pi i Nt}\sum_{\gamma \in \Z^n} 
e^{ 2 \pi i N [  \frac{\tau}{2} 
\langle -\xi + \frac{\mu}{N}+ \gamma, -\xi +\frac{\mu}{N} + \gamma \rangle + \langle 
\frac{\mu}{N} -\xi +\gamma, x\rangle]} 
\leqno(5.8.9\mu)$$
as stated in (5.7.1).

Consider in particular the case of dimension n=1.  Then
 relative to the basic theta functions $\vartheta^{\tau}_{\mu,N}$ 
the elements $V = (0, \frac{1}{N}, 1)$ and 
$U = (\frac{1}{N},0,1)$ of $Heis(\Z/N)$ are representated by the matrices
$$V: e_1 \rightarrow e_2,\dots, e_n \rightarrow e_1\;\;\;\;\;\;
U:= diag(1, e^{2\pi i \frac{1}{N}},\dots, e^{2\pi i (N-1)\frac{1}{N}})$$
where $\{e_i\}$ denotes the standard basis of $\C^N$. 
These elements satisfy $UV = e^{\frac{2\pi i}{N}}VU$ hence generate the rational
rotation algebra ${\cal M}_{\frac{1}{N}}$
 with Planck constant $h = \frac{1}{N}$.  Hence $\Theta_N$ determines
a finite dimensional representation $\pi$ of this algebra, with image the group algebra
$\C[Heis(\Z/N)]$ of the finite Heisenberg group.  Moreover, the 
transformation laws define $Sp_{\theta}(2,
\Z/N)$ as a covariant group of automorphisms of $\C[Heis(\Z/N)]$.  From the dynamical
point of view, these automorphisms are very different from the automorphisms
defined by $Sp_{\theta}(2,\Z/N)$
on ${\cal M}_{\frac{1}{N}}$ (as in [B][B.N.S]): Indeed, 
 the representation $\pi$ kills the center of ${\cal M}_{\frac{1}{N}}$, and since it
is finite dimensional representation the automorphisms have discrete spectra.

\medskip

\subsection*{(5.9) Finite degree Cauchy-Szego projectors and change of complex structure}
\medskip

As mentioned several times above, we would like to view the transformation laws
as defining a unitary operator on the space $\Theta^i_N$ of theta functions with
a fixed complex structure.  However, as things stand,
the transformation laws (5.7.3) change the complex structure $\tau$
into $\tau' = \frac{a\tau - b}{-c + d\tau}$.  The purpose of this section is to
use the degree N Cauchy-Szego projector to change the complex structure back to
$\tau$.

It is right at this point that the Toeplitz method differs most
markedly from the Kahler quantization method of [A.dP.W] [We].  In the Kahler scheme,
the unitary (BKS)  operator  carrying $\tilde{Th}_N^{\tau'}$
back to  $\tilde{Th}_N^{\tau}$ is  parallel translation with respect to a natural flat
connection on the vector bundle $\tilde{\Theta}_N$ over the moduli space of complex
structures, whose fiber over $\tau$ is the space  $\tilde{Th}_N^{\tau}$.
 As discussed in these articles, the
connection is defined by the heat equation for theta functions.  Since  the {\it classical
theta functions} are solutions of this equation, they are already a parallel family
with respect to the connection--hence the unitary BKS operator in the Kahler setting
  is simply to 'forget'  the
change in complex structure $\tau \rightarrow \tau'$. 
Thus, the unitary matrix defined relative to the classical theta functions
is precisely the quantization of $g$ in the Kahler sense.  It is also
the quantization of  [B.H][dB.B][dE.G.I][Ke], as the interested reader may confirm by 
comparing their formulae for the quantized cat maps with the expressions in the
transformation formulae.  

Our purpose now is to show that the Toeplitz method leads to the same result. 
\medskip

\subsection*{(5.9.1) Lemma} {\it Let  $\Pi^{\tau}_N$ be the orthogonal projection
onto $H^2_{\Sigma_{\tau}}(N)\equiv \Theta_N^{\tau}$ and let 
$\Pi_{N}^{\tau,\tau'}:= \Pi_N^{\tau} \Pi_N^{\tau'}: \Theta_N^{\tau'} \rightarrow \Theta_N^{\tau}.$ 
   Then: $\Pi_N^{\tau,\tau '*} \Pi_N^{\tau,\tau'} = 
(4\pi)^{ n}|\frac{( Im \tau Im \tau')^{\frac{n}{4}}}
{(-2\pi i (\tau-\overline{\tau'}))^{\frac {n}{2}}}|^2
 \Pi_N^{\tau}.$} 

\subsection*{Proof}: 

 Let $f
\in \Theta_N^{\tau}$, and  $g\in \Theta_N^{\tau'}$,
for any pair of complex structure $\tau,\tau'$.  As elements
of $L^2(N_{\Z}/N_{\R})$ their inner product is given by
$$(f|g):= \int_{N_{\Z}\backslash N_{\R}} (f|_n) \overline{(g|_n)}dn$$
with  $dn = dx d\xi dt$  the  $N_{\R}$-invariant measure on $N_{\Z}\backslash N_{\R}.$
Our main task is to calculate the inner products
$$(\vartheta^{\tau}_{\mu,N}| \vartheta^{\tau'}_{\mu',N})$$ in
$H_N (N_{\Z}\backslash N_{\R})$.
The Lemma is  equivalent to the following
\medskip

\noindent{\bf (5.9.2)  Claim}:
$$(\vartheta^{\tau}_{\mu, N} \;|\; \vartheta^{ \tau'}_{\mu', N}) =
\delta_{\mu,\mu'} vol(\R^n/\Z^n) (-2\pi i N(\tau- \overline{\tau'}))^{-\half n}. $$

\subsection*{Proof of Claim}

Using the expressions in (5.7.1)-(5.8.9$\mu$), we can rewrite the inner product in the form
$$(\vartheta^{\tau}_{\mu, N} \;|\; \vartheta^{ \tau'}_{\mu', N})=
\sum_{\gamma,\gamma' \in \Z^n}
\int_{\R^n/\Z^n} \int_{\R^n/\Z^n} e^{2\pi i N [\half \tau \langle  \gamma + \frac{\mu}{N}+\xi, 
\gamma + \frac{\mu}{N}+\xi\ \rangle
 +\langle \gamma + \frac{\mu}{n}+\xi,x\rangle ] }\;\;\;\leqno(5.9.3)$$
$$\;\;\;\;\;\;\;\;\;\;\;\;\;\;\;\;\;\;\;\;\;\;\;\;\;\;\;\;\;
 e^{-2\pi i N [\half\overline{\tau'} \langle \gamma' + \frac{\mu'}{N}+\xi, 
\gamma' + \frac{\mu'}{N}+\xi\rangle +\langle  
\gamma' + \frac{\mu'}{N}+\xi,x\rangle ]}\;\;\;dx d\xi.$$
The $dx$ integral equals
$$ \int_{\R^n/\Z^n}  e^{2\pi i  \langle x, N(\gamma -\gamma' )+ (\mu -\mu')\rangle }dx =
\delta_{N\gamma + \mu,N\gamma' + \mu'}.$$
 Since 
$$\delta_{N\gamma + \mu,N\gamma' + \mu'} = \delta_{\Z^n + \frac{\mu}{N},
 \Z^n + \frac{\mu'}{N}} \delta_{\gamma,\gamma'}$$
the expression in (5.9.3) simplifies to
$$\sum_{\gamma}\int_{\R^n/\Z^n}  e^{2\pi i N\{\half(\tau-\overline{\tau'})\}
 \langle \gamma + \frac{\mu}{N}+\xi, 
\gamma + \frac{\mu}{N}+\xi\rangle } d\xi
=\int_{\R^n}  e^{2\pi i N \{\half(\tau-\overline{\tau'})\}\langle \xi, \xi\rangle } d\xi.$$
The last expression is an inner product of Gaussians, so by (5.3.9) it equals
$$ \langle \gamma_{\tau}, \gamma_{\tau'}\rangle 
= (-2\pi i N(\tau- \overline{\tau'}))^{-\half n}\leqno(5.9.4)$$
proving the Claim.

It follows first that for each $\tau$  the basis $\{\vartheta^{\tau}_{\mu,N}, 
\mu \in \Z^n / N\Z^n\}$
is orthonormal up to the factor $(4\pi N  Im \tau)^{-\half n}$.  If we normalize
the basis to $\tilde{\vartheta}^{\tau}_{\mu,N}:=( 4\pi N Im \tau)^{\frac{1}{4} n}
\vartheta^{\tau}_{\mu,n}$ then the projection $\Pi_N^{\tau}$ onto the space
 $H_{\Sigma_{\tau}}^{2}(N)$
of degree N $\vartheta^{\tau}$'s  may be written in the form
$$\Pi_N^{\tau} = \sum_{\mu \in \Z^n/N\Z^n}\tilde{\vartheta}^{\tau}_{\mu,N}
\otimes \tilde{\vartheta}^{\tau*}_{\mu,N}. \leqno(5.9.5)$$
We then have  
$$\Pi_N^{\tau, \tau'*}\Pi_N^{\tau, \tau'}=\Pi_N^{\tau}\Pi_N^{\tau'}\Pi_N^{\tau}$$
$$=\sum_{\mu \in \Z^n / N\Z^n} |(\tilde{\vartheta}^{\tau}_{\mu,N}|
\tilde{\vartheta}^{\tau'}_{\mu,N})|^2 \tilde{\vartheta}^{\tau}_{\mu,N}\otimes
\tilde{\vartheta}^{\tau*}_{\mu,N}
= (4\pi)^{ n}|\frac{( Im \tau Im \tau')^{\frac{n}{4}}}
{(-2\pi i (\tau-\overline{\tau'}))^{\frac {n}{2}}}|^2
\sum_{\mu} \tilde{\vartheta}^{\tau}_{\mu,N}
\otimes\tilde{\vartheta^*}^{\tau}_{\mu,N}$$
proving the Lemma.\qed
\medskip

\subsection*{(5.9.6) Corollary}{\it Let 
$${\cal U}_N^{\tau,\tau'}: H_{\Sigma_{\tau'}}^2(N) \rightarrow H_{\Sigma_{\tau}}^2(N)$$
be the unitary opeator 
$${\cal U}_N^{\tau,\tau'}:= \sum_{\mu \in \Z / N\Z}\tilde{\vartheta}^{\tau}_{\mu,N}
\otimes \tilde{\vartheta}^{\tau'*}_{\mu,N}.$$
Then $$\Pi^{\tau,\tau'}_N =
(4 \pi)^{\half n}
 [\frac{( Im \tau Im \tau')^{\frac{n}{4}}}{(-2\pi i (\tau-\overline{\tau'}))^{\frac {n}{2}}}]
 {\cal U}_N^{\tau,\tau'}.$$}

We can now complete the
\medskip

\noindent{\bf (5.10) Proof of Theorem D}.
\medskip

\noindent(a)  Since $\Pi T_{\chi_g}\Pi$ is block diagonal relative to the decomposition (5.2.6)
it suffices to show that each block
 $\Pi_N^{i} T_{\chi_g} \Pi_N^{i}$ 
 is unitary up to a constant independent
of N. 

 For simplicity of notation let us rewrite the unitary coefficients under the
sum in (5.7.3) as $u_{\mu, \alpha}(g,N).$  Let us also observe that the norm of
$||\theta^{\tau}_{\mu,N}||$ varies with $\tau.$  Hence the transformation laws
(5.7.3) take the following form in terms of the $\tilde{\vartheta}^{ \tau}_{\mu,N}$'s:
$$\Pi^i T_{\chi_g} \Pi^i \tilde{\vartheta}^{ i}_{\mu,N}= j_g(g^{-1}\cdot i)^n
\frac{||\vartheta^{ g^{-1}\cdot i}_{\mu,N}||}{||\vartheta^{ i}_{\mu,N}||}     \nu(g, N) 
 \Pi^i \sum_{{\alpha \in \Z^n}
\atop{c \alpha mod N\Z^n}} u_{\mu, \alpha}(g,N) \tilde{\vartheta}^{ g^{-1}\cdot i}_{a\mu - 
c\alpha, N}.\leqno(5.10.1)$$
Using Corollary (5.9.6) and simplifying, (5.10.1) becomes 
$$\langle \gamma_{g^{-1}\cdot i}, \gamma_i \rangle 
  j_g(g^{-1}\cdot i)^n \nu(g, N){\cal U}_N^{i,g^{-1}i} \sum_{{\alpha \in \Z^n}
\atop{c \alpha mod N\Z^n}} u_{\mu, \alpha}(g,N) \tilde{\vartheta}^{ g^{-1}\cdot i}_{a\mu - 
c\alpha, N}$$
$$= \langle \gamma_{g^{-1}\cdot i}, \gamma_i \rangle
 j_g(g^{-1}\cdot i)^n  \nu(g, N) \sum_{{\alpha \in \Z^n}
\atop{c \alpha mod N\Z^n}} u_{\mu, \alpha}(g,N) \tilde{\vartheta}^{ i}_{a\mu - 
c\alpha, N}.\leqno(5.10.2)$$
Noting that $j_g(g^{-1}\cdot i) = \mu(g^{-1}, i)$ and comparing with Proposition
(5.3.8) we see that 
$$\Pi^i T_{\chi_g} \Pi^i  = \langle \mu(g^{*})e_{\Lambda},
e_{\Lambda} \rangle U_{g,N} \leqno (5.10.3)$$
where
$$U_{g,N} \tilde{\vartheta}^{ i}_{\mu,N} := \nu(g, N) \sum_{{\alpha \in \Z^n}
\atop{c \alpha mod N\Z^n}} u_{\mu, \alpha}(g,N) \tilde{\vartheta}^{ i}_{a\mu - 
c\alpha, N}. \leqno(5.10.4)$$
Hence by Corollary (5.3.10) we have
$$U_{g,N} = m(g) \Pi^i T_{\chi_g} \Pi^i $$
with $m(g) = \langle \mu(g^*)  e_{\lambda}, e_{\Lambda} \rangle^{-1} =
2^{-\frac{n}{2}} (det( A + D + iB - iC))^{\half}.$\qed

\medskip

 Comparing (5.10.2) and Corollary (5.3.10) we see that the principal symbol
is indeed the complete symbol.   
\medskip

\noindent(b) It is a classical fact that the transformation laws define the metaplectic 
representation of $SL_{\theta}(2, \Z/N)$.  We have defined the multiplier $m(g)$
precisely to obtain this representation.
\medskip

\noindent{\bf Remark} In the case of the real metaplectic representation, Daubechies [D]
finds that $W_J(S) = \eta_{J,S} P_J U_S |_{{\cal H}_J}$, where: $W_J(S)$ denotes
 the metaplectic representation, realized on the Bargmann space ${\cal H}_J$
of $J$-holomorphic functions;
 $U_S$ denotes left translation by $S^{-1}$, $P_J$; $P_J$ denotes the orthogonal onto
${\cal H}_J$; and  $\eta_{J,S}:= (\Omega_J, W_J(S) \Omega_J)^{*-1}$ [D., p.1388].
It is evident that in our notation $g = S^{-1}$ and that  $ m(g) =\eta_{J,S} $, corroborating
that  $m(g)$ is the correct multiplier to get the metaplectic representation.

\medskip

\noindent(c)  The index of $\chi_g$ is by definition the index of any Toepltiz Fourier
Integral operator $\Pi A T_{\chi_g}\Pi$ quantizing $\chi_g$ with unitary principal symbol.
We have seen that $m(g) \Pi T_{\chi} \Pi$ has a unitary principal symbol, and by (a) it
is actually a unitary operator.   Hence its index is zero. \qed
\medskip

\noindent(d) The ergodicity and mixing statements
 follow from Theorem B together with  the fact that symplectic torus automorphisms
are mixing if no eigenvalue is a root of unity [W]. 
\medskip

\noindent(e) We have:
$$U_g^* \Pi \sigma \Pi U_g = \Pi T_{\chi_g}^* \Pi \sigma \Pi T_{\chi_g} \Pi$$
as the remaining constant factors cancel.  The formula in (e) follows since
 $T_{\chi_g}^* \Pi T_{\chi_g}$ is
precisely the Toeplitz structure corresponding to the complex structure $g\cdot i$.
It also follows that the matrix elements of a Toeplitz operator relative to the
eigenfunctions $\vartheta^{i}_{k,N}$ of $U_{g,N}$ satisfy:
$$\langle \Pi \sigma \Pi \vartheta^{i}_{k,N}| \vartheta^{i}_{k,N}\rangle 
= \langle U_{g,N} \Pi \sigma \Pi \vartheta^{i}_{k,N}| U_{g,N}\vartheta^{i}_{k,N}\rangle
=\langle \Pi \sigma\cdot \chi_g \Pi \vartheta^{gi}_{k,N}| \theta^{gi}_{k,N}\rangle$$
where $\vartheta^{gi}_{k,N} = {\cal U}^{i,gi}_N \theta^{i}_{k,N}.$ \qed
\medskip

\noindent{\bf Remarks}

\noindent{\bf 1. On the index problem}  Weinstein's index problem actually
concerns Fourier Integral operators quantizing homogeneous canonical transformations
on $T^*M$ [Wei].  Of course, such a transformation is the same as a contact
transformation on
$S^*M$.  Moreover, it is known that any FIO can be expressed in the form $\Pi A
T_{\chi} \Pi$ where $\Pi$ is a Toeplitz structure on the symplectic cone generated by
the canonical contact form on $S^*M$ in $T^*(S^*M)$ and where $A$ is a
pseudodifferential operator on $S^*M$.
  Thus $\Pi$ is a Szego projector to a space $H^2(S^*M)$ of
CR functions on $S^*M$.  The Boutet de Monvel index theorem for pseudodifferential
Toeplitz operators and the logarithm law for the index reduce Weinstein's index
problem to that of calculating indices of operators of the form $\Pi T_{\chi}\Pi.$ It
is possible that the index of such an operator always vanishes; we have just seen
a non-trivial example of this (i.e. an example not homotopic to the identity thru
contact transformations).

The fact that the index vanishes for the symplectic torus automorphisms above is
due to the fact that their quantizations commute with an elliptic
circle action. Hence they are direct sums of finite rank operators and the index,
being the sum of indices of finite rank operators, has to vanish.  
It would be interesting to see if  the seemingly more difficult
index problem for Zoll surfaces (the original problem in [Wei]) cannot be solved
by a similar argument.  The main difference is that the contact map arising
there intertwines two different elliptic circle actions.

\noindent{\bf 2. On the quantum ergodicity} The quantum ergodicity theorem for
cat maps of $\R^2/\Z^2$ has previously been proved in [d'E.G.I] and [B.dB] by
a different method.  These papers also allow for non-trivial characters of the
fundamental group.

\bigskip

\section{Trace formulae for quantized torus automorphisms}
\bigskip

The purpose of this section is to prove an exact trace formula for the trace
$Tr U_{g,N}$ of a quantized cat map in the space of theta functions $\Theta_N$ of degree N.
The standard complex structure $\tau = iI$ is fixed throughout.  In the following we
assume that $g$ is non-degenerate in the sense that $ker(I-g)$ is trivial. The
square root $\sqrt{det(I-g)}$ is defined by the usual analytic continuation [F].

.

\noindent{\bf (6.1) Theorem E}{\it $\;\;$With the notations and assumptions of Theorem D,
and with the assumption that $g$ is non-degenerate,
we have:
$$Tr U_{g,N} =  \frac{ 1}{\sqrt{det(I-g)}}
 \sum_{[(m,n)] \in \Z^{2n}/ (I-g)^{-1}\Z^{2n}}
 e^{i \pi N [\langle m,n \rangle - \sigma ((m,n), (I-g)^{-1} (m,n))]}$$}
\medskip

\noindent{\bf Proof}: 

Our starting point is the explicit form of the Szego kernel $S$ from $L^2(N_{\R})$
 on $N_{\R}$ (cf. [S]).  It is a convolution kernel
 $S(x,y)= K(x^{-1}y)$ with
$$K(x) = c_n \partial_t (t + i|\zeta|^2)^{-n} \leqno(6.2)$$
where $c_n$ is a constant whose value we will not need to know, and where
  $x = (\zeta,t)$.  The Szego kernel $S(x,y)$ is singular along the diagonal, but
it can be regularized in a well-known way (see [S]) and we can safely pretend that it is
regular.  In fact, we will not need the full Szego kernel, but only the part of degree N,
and this is regular.

The kernel of  $\Pi T_{\chi_g} \Pi$ on $N_{\R}$ is then given by
$S(x, g(y))= c_n K(x^{-1}g(y))$.  Since $g$ is an automorphism, the kernel  on the quotient is
$$c_n \sum_{\tau \in N_{\Z}} K(x^{-1} \tau g(y)).$$
Actually, it
 will prove convenient to put the quotient kernel in a slightly different form by passing to
the quotient in two stages.  First, we sum over the central lattice $N_{\Z} \cap Z_{N_{\R}}$
to get the kernel of the Szego projector on the reduced Heisenberg group $\Hb^{red}$: 
$$S_{red}(x, y):= \sum_{k \in \Z} S(x, (0,0,k) y).$$
Since the part of degree N on $\Hb^{red}$ is  given by
$$S_N (x,y) = \int_o^1 S_{red}(x, y(0,0,\theta)) e^{- 2\pi i N \theta} d\theta.\leqno(6.3)$$
we may express it in the form
$$S_N(x, y)= \int_{\R} S(x, y(0,0,\theta)) e^{-2\pi i N \theta} d\theta.$$
The degree N part 
of $\Pi T_g \Pi$
 on $\Hb^{red}$ is therefore given by
$$S_N(x, g(y))= \int_{\R} S(x, g(y)(0,0,\theta)) e^{-2\pi i N \theta} d\theta.\leqno(6.4)$$

To pass to the full quotient we must further divide by the covering group $\bar \Gamma$ of
$\Hb^{red}_n$ over $N_{\R}/ N_{\Z}$.  It is not quite $\Z^{2n}$ since
the latter is not a subgroup of the Heisenberg group.  Rather  $\Z^{2n}$ is a maximal isotropic
subgroup of $K = \R^{2n}$ and we must embed it in $\Hb^{red}_n$
  by the splitting homomorphism
$$s: \Z^{2n} \rightarrow \Hb^{red}_n\;\;\;\;\;\; s(m,n) =(m,n, e^{i\half F(m,n)}) $$ 
with  $F(x,y) = \langle x, y \rangle$.
(cf. \S5.8).

  Since $g$ is an automorphism of the reduced Heisenberg group,
the kernel of the degree N part of $\Pi T_{\chi_g} \Pi$ on the full quotient can then
 be expressed  in the form:
$$\Pi_N T_{\chi_g}\Pi_N = 
c_n \sum_{\gamma \in\bar\Gamma}\int_{\R}  K(x^{-1} \gamma g(y) (0,0,\theta))
e^{-2\pi i N \theta} d\theta. \leqno(6.5N)$$

 Now denote a fundamental domain for $N_{\Z}$ in $N_{\R}$
by ${\cal D}$.  Then we have:
$$Tr \Pi_N T_{\chi_g} \Pi_N = c_n \sum_{\gamma \in \bar \Gamma}\int_{\R} 
\int_{{\cal D}} K(x^{-1}\gamma g(x)(0,0,\theta))
e^{-2\pi iN \theta} d\theta dx.\leqno(6.6N)$$
To simplify (6.6N), we 
 define an equivalence relation on $\bar \Gamma$:
$$\gamma \sim \gamma' \equiv \exists M \in \Gamma:\leqno(6.7)$$
$$\gamma' = M^{-1} \gamma g(M).$$
Here $g(M)$ denotes the  value of $g \in Sp(2n,\Z)$ on $M$ in $\Hb^{red}_n$.
  We denote the set of equivalence classes $[\gamma]$ by $[\bar \Gamma]$ . 

It follows from (6.6N), and  (6.7)  that the trace may be re-written in the form:
$$Tr \Pi_N T_{\chi_g} \Pi_N = c_n \sum_{[\gamma]} \sum_{M \in\bar \Gamma} \int_{{\cal D}} \int_{\R}
K(x^{-1}M^{-1}[\gamma] g(M) g(x)(0,0,\theta)) 
 e^{-2\pi i N \theta} d\theta dx .\leqno(6.8N)$$
We now use that $g$ is an automorphism to rewrite $g(M) g(x)$ as $g(Mx).$
Changing variables to $x'= Mx$ and noting that $\bigcup M{\cal D} = \R^{2n}\times S^1$,
we have:
$$Tr \Pi_N T_{\chi_g} \Pi_N = c_n  \sum_{[\gamma] \in [\bar \Gamma]} \int_{\R^{2n}\times S^1}\int_{\R}
K(x^{-1}[\gamma]g(x) (0,0,\theta)) e^{-2 \pi i N \theta} d\theta dx \leqno(6.9N).$$
We now observe that the central part of $x$ cancels out, so that we may replace
the reduced Heisenberg group by $\R^{2n}$.  We henceforth denote
points in this space by $\zeta= (x, \xi)$.  

 Since
$s :\Z^{2n} \rightarrow\bar \Gamma$ is an isomorphism, the equivalence classes
$[\gamma]$ in $\bar \Gamma$ are
in 1-1 correspondence with the cosets $[m,n]$ in $\Z^{2n}/ (g-I) \Z^{2n}.$  We denote
the latter set of equivalence classes by $[\Z^{2n}]$ and rewrite
 (6.9N) in the form
$$c_n  \sum_{[(m,n)] \in [\Z^{2n}]} \int_{\R^{2n}}\int_{\R}
K((-\zeta,0)(m,n,\half F(m,n))(g \zeta,0) (0,0,\theta))e^{-2\pi i N\theta}
 d\theta dx d\xi.\leqno(6.10N)$$
 We now multiply out the argument in $K$.  Since it is somewhat more convenient,
we express the result for the reduced form of the Heisenberg group:
$$(-\zeta,1) ((m,n),e^{i\pi F(m,n)}) (g \zeta,1) =
((g - I) \zeta + (m,n), e^{i \pi 
 \sigma ((m,n)-\zeta, g\zeta)}e^{-i\pi \sigma(\zeta, (m,n))}e^{i\pi F(m,n)}). \leqno(6.11)$$

Then write $(m,n)= (g-I)v$ and change variables $\zeta \rightarrow \zeta + v.$ Then (6.10N)
becomes
$$c_n \sum_{[(m,n)] \in [\Z^{2n}]}e^{i\pi N F(m,n)} \int_{\R^{2n}}\int_{\R}
K((g-I)\zeta, 0)(0,0,\theta)) e^{i \pi N \sigma(g(\zeta - v), \zeta - v)}
e^{i \pi N \sigma( (g-I)v, (g+I)[\zeta - v])} e^{-2\pi i N \theta} d\theta d\zeta. \leqno(6.12N)$$

Next we recall (cf. [S]) that the Fourier transform $\hat{K}$ as a function on $\R^{2n + 1}$ is
given by
$$\hat{K}(u,v,\tau) = 2^n e^{- \pi \frac{|(u,v)|^2}{2 \tau}} \;\;\;\;\;\;(\tau > 0).$$
Hence the partial Fourier transform in the $\theta$-variable equals
$$\hat{K}_{\theta}(\zeta, N) = 2^n c'_n N^{n} e^{-\pi N |\zeta|^2}$$
for another constant $c'_n$.
Therefore, (6.12 N) has the form
$$2^n c''_n N^{n} \sum_{[(m,n)] \in [\Z^{2n}]}  e^{i \pi N F(m,n)} \int_{\R^{2n}}
e^{-\pi  N |(g-I)\zeta|^2} 
 e^{i \pi N \sigma(g(\zeta - v), \zeta - v)}
e^{i \pi N \sigma( (g-I)v, (g+I)[\zeta - v])}   d\zeta\leqno(6.13N).$$
This is a Gaussian integral, and hence can be explicitly evaluated.  To do so, we first
simplify the  exponent.

First, the quadratic terms in $\zeta$ in the exponent are:
$$-\pi N [|(g-I)\zeta|^2 - i \sigma(g \zeta, \zeta)]$$
while the linear terms are:
$$i\pi N [\sigma( g\zeta, -v) + \sigma(-gv, \zeta) + \sigma((g-I)v, (g+I)\zeta).$$
The terms independent of $\zeta$ come to:
$$i \pi N [ \sigma(gv,v) + \sigma((g-I)v, (g+I)(-v)) + F(m,n)].$$
which simplify to
$$i \pi N [ F(m,n) - \sigma ((m,n),v)]$$
since $g$ is symplectic.  The terms linear in $\zeta$ cancel out.  

Hence,
$$Tr \Pi_N T_{\chi_g} \Pi_N =  2^n c''_n N^{n} I_{g,N} \sum_{[(m,n)] 
\in [\Z^{2n}]} e^{i \pi N [F(m,n) - \sigma ((m,n),v)]} \leqno(6.14N)$$
with
$$I_{g,N} = \int_{\R^{2n}} e^{-\pi N |(g-I)\zeta|^2} e^{i \pi N \sigma(g \zeta, \zeta)} d\zeta.$$
This integral has been evaluated in [D, p. 1386] and equals
$$N^{- n} c'''_n [det (I - g- i J(I + g))]^{-\half} [det(I - g)]^{-\half} \leqno(6.15)$$
for some normalizing factor  $c'''_n$.  It follows that 
$$Tr \Pi_N T_{\chi_g} \Pi_N =  2^n C_n  [det (I - g- i J(I + g))]^{-\half}
 [det(I - g)]^{-\half} \sum_{[(m,n)] 
\in [\Z^{2n}]} e^{i \pi N [F(m,n) - \sigma ((m,n),v)]}\leqno(6.16N)$$
for some constant $C_n$.  Using the remark after Theorem D(b) and using the formula
$$2^n  [det (I - g- i J(I + g))]^{-\half} = m(g)^{-1} \leqno (6.17)$$
from [D, p.1388] we see that
$$Tr U_{g,N} = C_n  [det(I - g)]^{-\half} \sum_{[(m,n)] 
\in [\Z^{2n}]} e^{i \pi N [F(m,n) - \sigma ((m,n),v)]}\leqno(6.18N)$$
for some constant $C_n$.
 We can determine this constant by computing one non-degenerate
example; the example we choose is the finite Fourier transform $F(N)$, whose trace
is given after the statement of Theorem E in \S 1.  Comparing with (6.18N) we find
that $C_n = 1.$ \qed

 \addtolength{\baselineskip}{-4pt}


\begin{thebibliography}{ABCD}


\bibitem[At]{At} M. Atiyah, {\it The Geometry and Physics of Knots}, Lezioni Lincee,
Cambridge Univ. Press (1990).

\bibitem[A.T]{A.T} L.Auslander and R.Tolimieri, Is computing with the finite Fourier
transform pure or applied mathematics, Bull.AMS 1 (1979), 847-897.

\bibitem[A]{A} L. Auslander, {\it Lecture Notes on
Nil-theta Functions}, CBMS series no.\ 34, AMS
Publications (1977). 

\bibitem[A.dP.W]{A.dP.W}  S.Axelrod, S.della Pietra, and E.Witten, Geometric quantization of
the Chern-Simons gauge theory, J.D.G. 33 (1991) 787-902.

\bibitem[Bai]{Bai} W.Baily, Classical theory of $\theta$-functions, in AMS Proc.Symp.Pure.
Math. IX, AMS (1966), 306-311.

\bibitem[B]{B} F.Benatti, {\it Deterministic Chaos in Infinite Quantum Systems}, Trieste
Notes in Physics, Springer-Verlag (1993).

\bibitem[B.N.S]{B.N.S}  F.Benatti, H.Narnhofer, and G.L.Sewell, A non-commutative version
of the Arnold cat map, Lett.Math.Phys. 21 (1991), 157-172.



\bibitem[B.P.U]  D.Borthwick, T.Paul, A.Uribe, On the non-vanishing of Poincare series
of large weight, to appear in Inv.Math.

\bibitem[B]{B} L. Boutet de Monvel, Toeplitz
operators---an asymptotic quantization of symplectic
cones, in:  {\it Stochastic Processes and Their
Applications}, S. Albeverio (Ed.), Kluwer Acad.\ Pub.\,
Netherlands (1990). 

\bibitem[B.G]{B.G} L. Boutet de Monvel and V. Guillemin,
{\it The Spectral Theory of Toeplitz Operators}, Ann.\
Math.\ Studies {\bf 99}, Princeton U. Press (1981). 

\bibitem[B.S]{B.S} L. Boutet de Monvel and J. Sj\"ostrand,
Sur la singularite des noyaux de Bergmann et de Szeg\"o,
Asterisque {\bf 34-35} (1976) 123-164.

\bibitem[B.dB]{B.dB} A.Bouzouina and S. de Bievre , Equipartition of the eigenfunctions
of quantized ergodic maps on the torus, to appear in Comm.Math.Phys.


\bibitem[B.R]{B.R} O. Bratteli and D.W. Robinson, {\it
Operator Algebras and Quantum Statistical Mechanics I},
Springer-Verlay (1979). 

\bibitem[C]{C} P.Cartier, Quantum mechanical commutation relations and theta
functions, in AMS Symposium ????

\bibitem[CV]{CV} Y. Colin de Verdiere, Ergodicit\'e et
functions propres du Laplacian, Comm.\ Math.\ Phys.\ {\bf 102} (1985),
497-502. 

\bibitem[D]{D} I. Daubechies, Coherent states and projective representations of the
linear canonical transformations, J.Math.Phys. 21 (1980), 1377-1389.



\bibitem[dE.G.I]{dE.G.I} M. d'Egli Esposti, S. Graffi, and S.
Isola, Stochastic properties of the quantum Arnold cat in
the classical limit, Comm.Math.Phys. 167 (1995), 471-509.

\bibitem[D]{D} R.G.Douglas, {\it $C^*$- Algebra Extensions and K-Homology}, Ann.Math.Studies
no.95, Princeton U.Press, Princeton (1980).

\bibitem[F]{F} G. Folland, {\it Harmonic Analysis in Phase
Space}, Ann.\ Math.\ Studies, no.\ {\bf 122}, Princeton U.
Press (1989). 

\bibitem[F.S]{F.S} G. Folland and E. Stein, Estimates for the $\bar
\partial_b$ complex and analysis on the Heisenberg group, Comm.\ P.A.M.
{\bf 27} (1974), 429-522. 

\bibitem[G.1]{G.1} V.Guillemin,  Residue traces for
certain algebras of Fourier Integral operators, J. Fun.\
Anal.\ {\bf 115} (1993), 381-417.

\bibitem[G.2]{G.2} V. Guillemin, A non-elementary proof of quadratic reciprocity (unpublished
manuscript).

\bibitem[H.B]{H.B}  J.H. Hannay and M.V. Berry, Quantization
of linear maps on a torus, Physica D1 (1980), p.\ 267. 


\bibitem[H]{H} E.J. Heller, In:  Chaos and Quantum Physics,
Les Houches 1989 (ed.\ by M.J. Giannoni, A. Voros and J.
Zinn-Justin), Amsterdam:  North Holland (1991).

\bibitem[Herm]{Herm} C.Hermite, Sur quelques formules relatives a la transformation
des fonctions elliptiques, Journal de Liouville III(1858), p.26.

\bibitem[J.P]{J.P} V.Jaksic and C.A.Pillet, Ergodic properties of classical dissipative
systems (preprint 1996).

\bibitem[K]{K} V.Kac, {\it Infinite Dimensional Lie Algebras}, 3rd ed.
 Cambridge: Cambridge Univ.Press (1990).

\bibitem[K.P]{K.P} V.Kac and D.H.Peterson,  Infinite dimensional Lie algebras,
theta functions and modular forms, Adv in Math. 53 (1984), 125-264.


\bibitem[Ke]{Ke} J. Keating, The cat maps:  quantum
mechanics and classical motion, Nonlinearity {\bf 4}
(1991), 309-341.

\bibitem[Kloo]{Kloo} H.D. Kloosterman, The behaviour of general theta functions under
the modular group and the characters of binary modular congruence groups.I, Ann.Math.
47(1946), p. 317.

\bibitem[M]{M} D.Mumford,{\it  Tata Lectures on Theta III}, Progress in Math. 97,
 Birkhauser, Boston (1991).

\bibitem[N.T1]{N.T1} H.Narnhofer and W.Thirring, Transitivity and ergodicity of
quantum systems, J.Stat.Phys. 52(1988), 1097-1112.

\bibitem[N.T.2]{N.T.2} -----------------------, Mixing properties of quantum
systems, J.Stat.Phys. 57 (1989), 811-825.

\bibitem[R]{R} D. Ruelle, {\it Statistical Mechanics},
Benjamin (1969).


\bibitem[Sn]{Sn} A.I. Snirelman, Ergodic properties of
eigenfunctions, Usp.\ Math.\ Nauk.\ {\bf 29} (1974),
181-182.

\bibitem[S]{S} E.Stein, {\it Harmonic Analysis}, Princeton: Princeton Univ.Press (1993).

\bibitem[Su]{Su} T. Sunada, Quantum ergodicity (preprint
1994).

\bibitem[Th]{Th} W.Thirring, {\it A Course in Mathematical Physics, vol. 4: Quantum
Mechanics of Large Systems}, Springer-Verlag, New York (1983).

\bibitem[U.Z]{U.Z} A.Uribe and S.Zelditch,  Spectral statistics on Zoll surfaces,
Comm.Math.Phys. 154 (1993), 313-346.

\bibitem[W]{W} P. Walters, {\it An Introduction to Ergodic
Theory}, Graduate Texts in Math.\ {\bf 79},
Springer--Verlag, NY (1982).

\bibitem[Wei]{Wei} A. Weinstein, Fourier Integral Operators, quantization,
and the spectrum of a Riemannian manifold, Colloques Internationaux
C.N.R.S. {\bf 237}, Geometrie Symplectique et Physique (1976).

\bibitem[We]{We} J.Weitsman, Quantization via real polarization of the moduli
space of flat connections and Chern-Simons gauge theory in genus one, Comm.Math.Phys.
137 (1991), 175-190.

\bibitem[Z.1]{Z.1} S.Zelditch,  Quantum ergodicity of $C^*$-dynamical systems
 (Comm.Math.Phys. 177 (1996), 507-528.

\bibitem[Z.2]{Z.2} ---------,  Quantum Mixing (J. Fun. Anal., to appear).

\bibitem[Z.3]{Z.3} \rule{.75in}{.005in}, Quantum transition
amplitudes for ergodic and for completely integrable
systems, J. Fun.\ Anal.\ {\bf 94} (1990), 415-436.


 \end{thebibliography}
\end{document}